\documentclass[preprint,prc,aps,showpacs,showkeys,groupedaddress,floatfix,natbib]{revtex4-1}
\usepackage{epsfig}
\usepackage{dcolumn}
\usepackage{bm}
\usepackage[latin5]{inputenc}
\usepackage{graphics}
\usepackage{graphicx}
\usepackage{epstopdf}
\usepackage{epsfig}
\usepackage{amssymb}
\usepackage{amsmath}
\usepackage{color}
\usepackage{natbib}

\newcommand{\soutb}{\bgroup\markoverwith{\textcolor{blue}{\rule[.5ex]{2pt}{1pt}}}\ULon}
\newcommand{\soutr}{\bgroup\markoverwith{\textcolor{red}{\rule[.5ex]{2pt}{1pt}}}\ULon}
\newcommand{\soutg}{\bgroup\markoverwith{\textcolor{green}{\rule[.5ex]{2pt}{1pt}}}\ULon}

\begin{document}
\title{Scattering of  Klein-Gordon particles in the background of mixed scalar-vector generalized symmetric Woods-Saxon potential}
\author{B.C. L\"{u}tf\"{u}o\u{g}lu}
\affiliation{Department of Physics, Akdeniz University, 07058 Antalya, Turkey}
\email{bclutfuoglu@akdeniz.edu.tr}
\author{J. Lipovsk\'{y}}
\affiliation{Department of Physics, Faculty of Science, University of Hradec Kr\'{a}lov\'{e}, Rokitansk\'{e}ho 62, 500\,03 Hradec Kr\'{a}lov\'{e}, Czechia}
\author{J. K\v{r}\'{i}\v{z}}
\affiliation{Department of Physics, Faculty of Science, University of Hradec Kr\'{a}lov\'{e}, Rokitansk\'{e}ho 62, 500\,03 Hradec Kr\'{a}lov\'{e}, Czechia}
\date{\today}
\begin{abstract}

Recently, it has been shown that the generalized symmetric Woods-Saxon potential energy, in which surface interaction terms are taken into account, describes the physical processes better than the standard form. Therefore in this study, we investigate the scattering of Klein-Gordon particles  in the presence of both generalized symmetric Woods-Saxon vector and scalar potential.  In one spatial dimension we obtain the solutions in terms of hypergeometric functions for spin symmetric or pseudo-spin symmetric cases. Finally, we plot transmission and reflection probabilities for incident particles with negative and positive energy for some critical arbitrary parameters and discuss the correlations for both cases.

\end{abstract}
 \maketitle 

\section{Introduction}
\label{intro}
Symmetry is a tool that we often consult to understand the nature. For instance in nuclear physics,   the pseudo-spin symmetry (PSS) and the spin symmetry (SS) are used to describe the nuclear structure phenomena \cite{RefBohr1982, RefMottelson1991, RefZhouMengRing2003}.  These symmetries were discovered in the first half of 1970's independently by Smith et al. \cite{RefSmithTassie1971}  and Bell et al. \cite{RefBellRuegg1975}. Since then, SS and PSS are being investigated in various scientific papers \cite{RefGinocchio1997, RefGinocchio1999, RefAlbertoCastroMalheiro2013, RefCastro2005, RefHamzaviRijabi2013, RefAlhaidariBahlouliAl-Hasan2006, RefAlbertoCastroMalheiro2007, RefAlbertoMalheiroFredericoCastro2015, RefIkotetal2015, RefAlbertoetal2016},  detailed reports have been published by Ginocchio \cite{RefGinocchio2005} and recently by Liang et al.  \cite{RefLiangMengZhou2015}. The origin of these symmetries basically depends on the investigated system's Hamiltonian, that is constituted with the external potential energy term limited with a Lorentz scalar, $V_s$, and a non-zero time component Lorentz vector, $V_v$, \cite{RefBlokhinBahriDraayer1995}. It is shown that SS arises in case of  $V_s-V_v=\varepsilon^+$ while $V_s+V_v=\varepsilon^-$ case results with PSS. Note that  $\varepsilon^+$ and $\varepsilon^-$ are constants and  they vanish for finite nuclei \cite{RefGinocchio1999}.

Alberto et al. \cite{RefAlbertoCastroMalheiro2007} gave the required conditions of the equivalent energy spectra of relativistic spin-half and spin-zero particles in the presence of vector and scalar potentials. All those conditions are independent of the potential parameters and emerge from the SS and the PSS in the Dirac equation.

In plenty of papers, the solutions of the Klein-Gordon (KG) and Dirac equation have been examined for different potential energies, without trying to name all, we mention e.g. Coloumb potential \cite{RefHamzavi2010, RefHassanabadiIkotYarrinkamar2014}, P\"oschl-Teller potential \cite{RefHassanabadiIkotYarrinkamar2014, RefIkotAtAll2015, RefHartmann2017}, Hulth\'en potential \cite{ RefIkotAtAll2016, RefGaoZhang2017}, Morse potential \cite{RefXieJia2015, RefZhanetal2016}. Among them as an example, Alhaidari et al.  \cite{RefAlhaidariBahlouliAl-Hasan2006} investigated the solutions  in three dimensions with vector and scalar potentials, which are non-central, i.e. depend also on other variables than the distance from the center, namely relativistic Coulomb, harmonic oscillator and Hartmann potential energies. They concluded that relativistic energy spectra are different from the well-known relativistic extensions, which have the same non-relativistic limit.

Apart from those investigations,  the famous  Woods-Saxon potential (WSP) has also been examined  \cite{ RefKennedy2002, RefPanella2010, RefOlgarMutaf2015}. Moreover Rojas et al. studied the KG particle scattering without scalar WSP \cite{RefRojasVIllalba} while Hassanabadi et al. with scalar potential \cite{HassanabadiMaghsoodiZarrinkamarSalehi2013}. Note that WSP first described by Woods and Saxon in \cite{RefWoodsSaxon1954} to describe the $20\,\mathrm{MeV}$ proton scattering from medium and heavy nuclei. The successive results that are compatible with the experiments motivate many scientists to investigate the nucleon interactions by using the WSP with the Coulomb potential energy too. The superposition of Coulomb and Woods-Saxon interaction can be described by Modified Woods-Saxon potential (MWSP) \cite{RefTianWangLi2007}. Recently in order to give a better description of  the energy barrier at the surface of atomic nucleus that nucleons suffer, Generalized Symmetric Woods-Saxon potential (GSWSP) was proposed \cite{CandemirBayrak2014, BayrakSahin2015, BayrakAciksoz2015, LutfuogluAkdeniz2016, RefLutfuoglu2018, LiendoCastro2016, BerkdemirBerkdemir2005, BadalovAhmado2009, GonulKoksal2007, KouraYamada2000, CapakPetrellis2015, CapakGonul2016, IkotAkpan2012, IkhdairFalayeHamzavi2013, surface1, surface2, surface3, surface4, surface5, RefKandirmaz2017}.

Since GSWSP takes the effects of surface interactions, in this manuscript we are motivated to examine the scattering of the KG particle for SS and PSS cases. The paper is structured as follows. In Section~\ref{sec:kg} we introduce KG equation and arrive at the time-independent KG equation for SS and PSS limit. The Section~\ref{sec:gsws} is devoted to solving both cases with the GSWS potential. We find solutions in the regions $x<0$ and $x>0$ as a combination of hypergeometric functions. Furthermore, we find forbidden region for the energy. Using the asymptotic behavior of the functions we obtain reflection and transmission coefficients. In Section~\ref{sec:res} we describe the behavior in the SS and PSS case for a neutral kaon scattering from a nucleus. This behavior can be seen from the dependence of the reflection and transmission coefficient on the energy plotted for several values of the coefficients determinating the potential. Finally, we conclude the results in Section~\ref{sec:conclusions}.

\section{Klein-Gordon equation with Vector and Scalar Coupling}\label{sec:kg}
The KG equation
\begin{eqnarray}
  \Big[\hat{p}^\mu \hat{p}_\mu - (m_0c)^2\Big]\Phi(\vec{r},t) &=& 0
\end{eqnarray}
represents  free spinless particle dynamics via two quantities, the rest mass $m_0$, and the linear momentum. Here, $\hat{p}^\mu$ denotes the four-momentum, while $c$ is the speed of light.

Electromagnetic interactions of KG particles can be introduced by
a gauge invariant coupling to the  four-vector linear momentum via the  minimal substitution
\begin{eqnarray}
  \hat{p}^\mu &\rightarrow & \hat{p}^\mu - \frac{e}{c} A^\mu,
\end{eqnarray}
here $e$ is a real parameter and represents the electromagnetic coupling term. $A^\mu$
is a gauge invariant four-vector potential that is constituted by time and space components.
Gauge invariance is very important since it leads to eliminating the nonphysical degrees of freedom.
Instead of choosing the Lorentz or Coulomb gauges, in this manuscript, we study a more simple problem by
taking the space component of the four-vector potential as zero while the time component is non-zero (i.e. $eA^0=V_v$).

The second parameter, the rest mass, is a scalar parameter
and can interact by using another coupling constant, $g$,
\begin{eqnarray}
  m_0 &\rightarrow & m_0 + \frac{g}{c^2}V_s,
\end{eqnarray}
where $V_s$ is a space-time scalar potential. In  strong regime, where $g\approx 1$, the scalar potential coupling  becomes important while in weak regime, $g \ll 1$, we immediately get the standard KG equation.

In this paper, we deal with $(1+1)$ Minkowski space-time. We assume that the wave function can be separated into space and time components, then from the time-dependent KG equation we arrive at the time-independent KG equation as follows
\begin{eqnarray}
  \Bigg[\frac{d^2 }{d x^2}+\frac{1}{\hbar^2c^2}\Big[ \big(E -V_v \big)^2- \big(m_0c^2 + g V_s\big)^2\Big]\Bigg]\phi(x) &=& 0, \,\,\,\,\,\,\,\,\,\,\,\,\,\,\,\,\,\,\,\, \label{KGham}
\end{eqnarray}
here $\hbar$ is the Planck constant. In the strong regime, the SS case is
\begin{eqnarray}
  V_v -  g V_s&=& \varepsilon^+,
\end{eqnarray}
and Eq.~(\ref{KGham}) yields to
\begin{eqnarray}
   \Bigg[\frac{d^2 }{d x^2}+\frac{1}{\hbar^2c^2} \bigg[E^2-\big(\varepsilon^+ -m_0c^2\big)^2 -2V_v\Big(E-\big(\varepsilon^+ -m_0c^2\big)\Big)\bigg] \Bigg]\phi(x) &=& 0.  \,\,\,\,\,\,\,\,\,\,\,\,\,\,\,\,
\end{eqnarray}
PSS solution
\begin{eqnarray}
  V_v +  g V_s&=& \varepsilon^-,
\end{eqnarray}
gives
\begin{eqnarray}
  \Bigg[\frac{d^2 }{d x^2}+\frac{1}{\hbar^2c^2} \bigg[E^2-\big(\varepsilon^- +m_0c^2\big)^2 -2V_v\Big(E-\big(\varepsilon^- +m_0c^2\big)\Big)\bigg] \Bigg]\phi(x) &=& 0.  \,\,\,\,\,\,\,\,\,\,\,\,\,\,\,\,
\end{eqnarray}
Here $\varepsilon^+$ and $\varepsilon^-$ are constants and for finite nuclei can be chosen as zero \cite{RefGinocchio1999}. In this paper, we investigate the SS and PSS cases on the same equation
\begin{eqnarray}
    \Bigg[\frac{d^2 }{d x^2}+\frac{1}{\hbar^2c^2}\Big[ (E^2-m_0^2c^4)-2(E\mp m_0c^2)V_v(x)\Big]\Bigg]\phi(x) &=& 0.
  \,\,\,\,\,\,\,\,\,\,\,\,\,\,\,\,\,\,\,\,\label{KG1}
\end{eqnarray}
Here $E\mp m_0c^2$ is used, $+$ indicates the SS while $-$  represents the PSS cases, respectively.

\section{Generalized Symmetric Woods-Saxon Potential}\label{sec:gsws}
The GSWS potential in the following form
\begin{eqnarray}
  V_v(x)&=&\theta{(-x)}\Bigg[\frac{-V_0}{1+e^{-a(x+L)}}+\frac{W e^{-a(x+L)}}{\big(1+e^{-a(x+L)}\big)^2}\Bigg] \nonumber \\
  &&+ \theta{(x)}\Bigg[\frac{-V_0}{1+e^{a(x-L)}}+\frac{W e^{a(x-L)}}{\big(1+e^{a(x-L)}\big)^2}\Bigg]. \label{WS1}
\end{eqnarray}
examined by \cite{LutfuogluAkdeniz2016} is employed. Here $\theta{(-x)}$ and $\theta{(x)}$ are the Heaviside step functions, $V_0$ represents the depth of the potential energy. $a$ and $L$ are  the reciprocal of the diffusion parameter and the radius that the potential energy is effective, respectively. GSWSP differs from WSP by  the second terms in the square brackets. These terms correspond to the energy barrier at the surface and it is linearly proportional to the spatial derivative of the first term times the effective radius. Thus, the parameter $W$ is linearly proportional to $a$, $L$, $V_0$ and a proportionality constant.  This new constant can be determined by means of momentum and energy conservations and does not have to be positive in general.  Note that in this manuscript all parameters are chosen as real and positive numbers. Moreover, in order to study a smooth potential energy function, parameters $a$ and $L$ are chosen in a way that their product is much greater than one.

\subsection{$x<0$ Region}
We start by substituting  the GSWSP given in Eq.~(\ref{WS1}) to the KG equation 
Eq.~(\ref{KG1}). In negative region we thus obtain
\begin{eqnarray}
  \Bigg[\frac{d^2}{dx^2}+a^2\Bigg(-\epsilon^2+\frac{\beta_\mp^2}{1+e^{-a(x+L)}}+\frac{\gamma_\mp^2}{\big(1+e^{-a(x+L)}\big)^2}\Bigg) \Bigg]\phi_L^{\mp}(x) &=& 0, \label{KG2x<0}
\end{eqnarray}
using the following new abbreviation parameters
\begin{eqnarray}
  -\epsilon^2&\equiv&\frac{(E^2-m_0^2c^4)}{a^2\hbar^2c^2}, \\
  \beta_\mp^2 &\equiv& \frac{2(E\mp m_0c^2)(V_0-W)}{a^2\hbar^2c^2},  \\
  \gamma_\mp^2 &\equiv& \frac{2(E\mp m_0c^2)W}{a^2\hbar^2c^2}.
\end{eqnarray}
Hereafter the encountered $\mp$ signs  indicate  the  solutions for the $V_v = \mp  g V_s $ cases. We use the transformation $ z \equiv \Big[1+e^{-a(x+L)}\Big]^{-1} $   to express KG equation as
\begin{eqnarray}
&&\Bigg[\frac{d^2}{dz^2}+\Big(\frac{1}{z}+\frac{1}{z-1}\Big)\frac{d}{dz}\nonumber
\\&&+\Bigg(\frac{\beta_\mp^2-2\epsilon^2}{z}-\frac{\epsilon^2}{z^2}
  -\frac{\beta_\mp^2-2\epsilon^2}{(z-1)}+\frac{\beta_\mp^2+\gamma_\mp^2-\epsilon^2}{(z-1)^2}\Bigg)\Bigg]  \phi_{L}^{\mp}(z) = 0. \label{KG4x<0}
\end{eqnarray}
The differential equation has two singular points, $z=0$ and $z=1$.  We need to study the dominant terms around these singular points to determine the behaviors of the solutions.

\begin{itemize}
  \item The dominant terms at $z=0$ singularity are  \begin{eqnarray}
    \Bigg[\frac{d^2}{dz^2}+\frac{1}{z}\frac{d}{dz}-\frac{\epsilon^2}{z^2}\Bigg]\phi_{L}^{\mp}(z) &\simeq& 0. \label{Sch4<0_1ax}
  \end{eqnarray}
We have a polynomial solution
\begin{eqnarray}
    \phi_{L}^{\mp}(z) &=& z^\mu\,,
\end{eqnarray}
where $\mu = \mp \big|  \epsilon\big|$ is chosen to be
\begin{eqnarray}
  \mu  &=& \frac{ik}{a}
\end{eqnarray}
with
\begin{eqnarray}
k&=&\frac{1}{\hbar c}\sqrt{(E\pm m_0c^2)(E\mp m_0c^2)}.\label{k1}
\end{eqnarray}

\item At $z=1$ singularity the dominant terms are  \begin{eqnarray}
    \Bigg[\frac{d^2}{dz^2}+\frac{1}{z-1}\frac{d}{dz} +\frac{\beta_\mp^2+\gamma_\mp^2-\epsilon^2}{(z-1)^2}\Bigg]\phi_{L}^{\mp}(z) &\simeq & 0. \label{Sch4<0_2ax}
  \end{eqnarray}
This is satisfied by a polynomial solution
\begin{eqnarray}
    \phi_{L}^{\mp}(z) &=& (z-1)^{\nu_\mp}\,,
\end{eqnarray}
where $\nu_\mp^2 + \beta_\mp^2+\gamma_\mp^2-\epsilon^2 = 0$
and
\begin{eqnarray}
  \nu_\mp&=& \frac{i\kappa_\mp}{a}
\end{eqnarray}
defined as
\begin{eqnarray}
  \kappa_\mp &\equiv& \frac{1}{\hbar c} \sqrt{(E\mp m_0c^2)(E\pm m_0c^2+2V_0)}\,. \label{k2}
\end{eqnarray}
\end{itemize}
Therefore the general solution of the differential Eq.~(\ref{KG4x<0}) could be given as the linear combination of both behaviors as
\begin{eqnarray}
  \phi_{L}^{\mp}(z) &=& z^\mu (z-1)^{\nu_\mp}f_\mp(z)
\end{eqnarray}
Then we find
\begin{eqnarray}
   &&z(1-z)f_\mp{''}(z)+ \Big[(1+2\mu)-(1+2\mu+2\nu_\mp+1)z\Big]f_\mp{'}(z)\nonumber \\&&-\Big[(\mu+ \nu_\mp)^2+(\mu+\nu_\mp)+\gamma_\mp^2\Big]f_\mp(z)= 0, \label{KG_eq5}
\end{eqnarray}
This result is a very well known form of the Hypergeometric differential equation
\begin{eqnarray}
   z(1-z)u{''}(z)+ \Big[c-(1+a+b)z\Big]u'(z)- ab u(z) &=& 0\label{Hypergeometric_dif_eq}
\end{eqnarray}
which has the following solutions
\begin{eqnarray}
  u(z) &=& A\,\,\, {}_2F_1[a,b,c;z]+B z^{1-c}\,\,\, {}_2F_1[1+a-c,1+b-c,2-c;z]\,, \label{hipergeometrik_genel_cozum}
\end{eqnarray}
where ${}_2F_1$ is a hypergeometric function. We compare the Eq.~(\ref{KG_eq5}) with Eq.~(\ref{Hypergeometric_dif_eq})
\begin{eqnarray}
  f_\mp(z) &=& D_1^\mp   \,\,\, {}_2F_1[\mu+\theta_\mp+\nu_\mp,1+\mu-\theta_\mp+\nu_\mp,1+2\mu;z]\nonumber \\
    &+& D_2^\mp z^{-2\mu}  \,\,\, {}_2F_1[-\mu+\theta_\mp+\nu_\mp,1-\mu-\theta_\mp+\nu_\mp,1-2\mu;z] \,, \,\,\,\,\,\,\,\,
\end{eqnarray}
where
\begin{eqnarray}
\theta_\mp \equiv \frac{1}{2}\mp \sqrt[]{\frac{1}{4}-\gamma_\mp^2}.
\end{eqnarray}
We write down the most general solution as
\begin{eqnarray}
  \phi_L^\mp(z) &=& D_1^\mp z^\mu (z-1)^{\nu_\mp} \,\,\, {}_2F_1[\mu+\theta_\mp+\nu_\mp,1+\mu-\theta_\mp+\nu_\mp,1+2\mu;z]\nonumber \\
  &+&D_2^\mp z^{-\mu} (z-1)^{\nu_\mp} \,\,\, {}_2F_1[-\mu+\theta_\mp+\nu_\mp,1-\mu-\theta_\mp+\nu_\mp,1-2\mu;z]. \,\,\,\,\,\,\,\,
\end{eqnarray}

\subsection{$x>0$ Region}
Since GSWSP energy is an even function, KG equation is covariant under the mapping $x\rightarrow -x$, therefore the general solution on the positive side will be  similar to the one at the  negative side as follows:
\begin{eqnarray}
  \phi_R^\mp(y) &=& D_3^\mp y^\mu (y-1)^{\nu_\mp} \,\,\, {}_2F_1[\mu+\theta_\mp+\nu_\mp,1+\mu-\theta_\mp+\nu_\mp,1+2\mu;y]\nonumber \\
  &+&D_4^\mp y^{-\mu} (y-1)^{\nu_\mp}\,\,\, {}_2F_1[-\mu+\theta_\mp+\nu_\mp,1-\mu-\theta_\mp+\nu_\mp,1-2\mu;y]. \,\,\,\,\,\,\,\,
\end{eqnarray}
Here,
$y \equiv \Big[1+e^{a(x-L)}\Big]^{-1} $
 coordinate transformation is used. Note that $D_i^\mp$, as $i=1, \dots, 4$ are normalization constants.

\subsection{Scattering  Conditions for Particles with Negative and Positive Energy}
Because of the existing symmetry, a relativistic spinless particle can scatter by approaching from minus or plus infinity. Here we assume that particle is moving from negative region to positive region. In order to describe particle scattering, two wave numbers defined by Eq.~(\ref{k1}) and Eq.~(\ref{k2}) should be real. This gives two conditions:
\begin{eqnarray}
(E+m_0c^2)(E-m_0c^2)&>&0 \label{eb1} \,,\\
(E\mp m_0c^2)(E\pm m_0c^2+2V_0) &>& 0\,. \label{eb2}
\end{eqnarray}
Eq.~(\ref{eb1}) holds true regardless the sign in  $V_v=\mp g V_s$ and implies that particles with positive energy, which have energies greater than the rest energy, can scatter whereas the particles with negative energy can be scattered only if they have energies smaller than $-m_0 c^2$.

Eq.~(\ref{eb2}) refers to two distinct sets of solutions as follows:

\subsubsection{Spin symmetric case}

\begin{figure*}
  \includegraphics[width=1.0\textwidth]{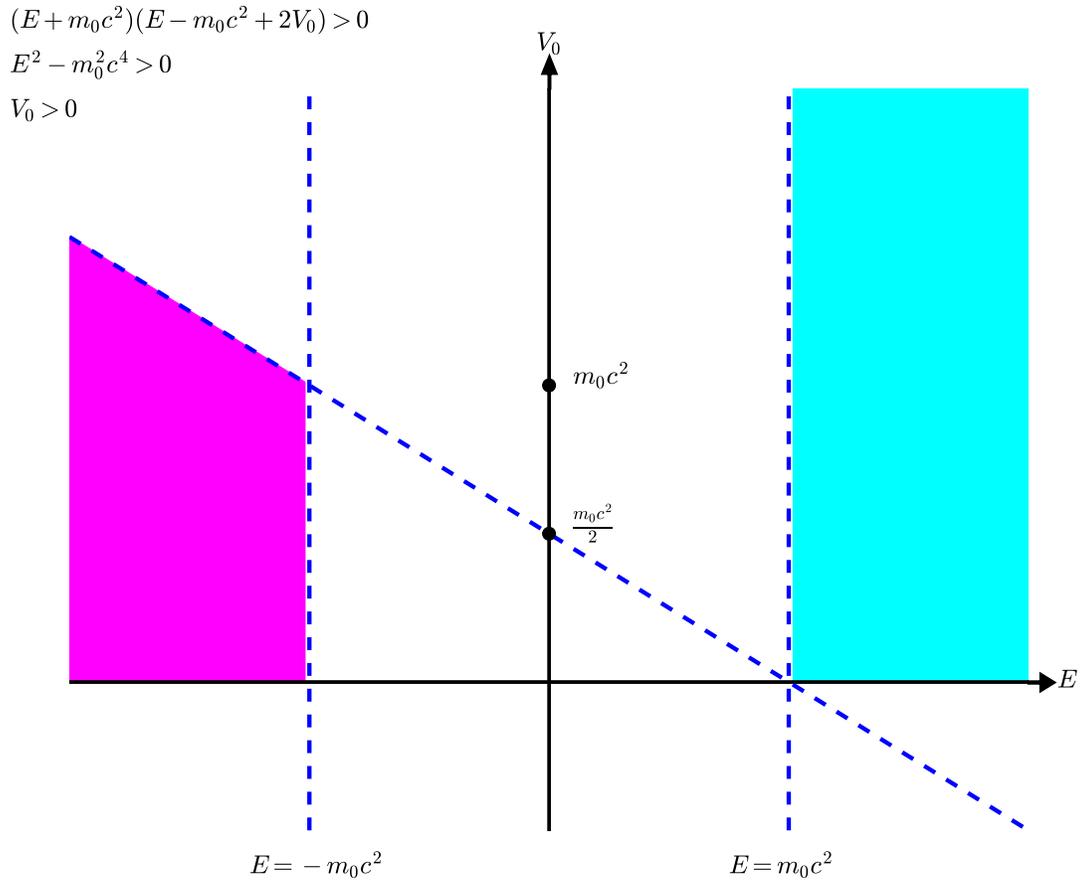}
\caption{Possible energy regions and forbidden energy gap for particle scattering in  SS case.}
\label{fig:1}       
\end{figure*}

The inequality in Eq.~(\ref{eb2}) imposes extra constraints on the forbidden gap in energy. To have a better understanding of the crucial role of the parameter $V_0$ we plot $E$ versus $V_0$  as given in Fig.~\ref{fig:1}. The incoming particles which have energies greater than $m_0c^2$ can scatter. This region is indicated by cyan color in Fig.~\ref{fig:1}. The incoming particles which have energies smaller than $-m_0c^2$ can only scatter if $V_0$ is smaller than $m_0 c^2$. When $V_0$ becomes greater than $m_0 c^2$, the scatterable negative energy region becomes bounded by $V_0$ as shown in the magenta region in  Fig.~{\ref{fig:1}}.

\subsubsection{Pseudo-spin symmetric case}

\begin{figure*}
  \includegraphics[width=1.0\textwidth]{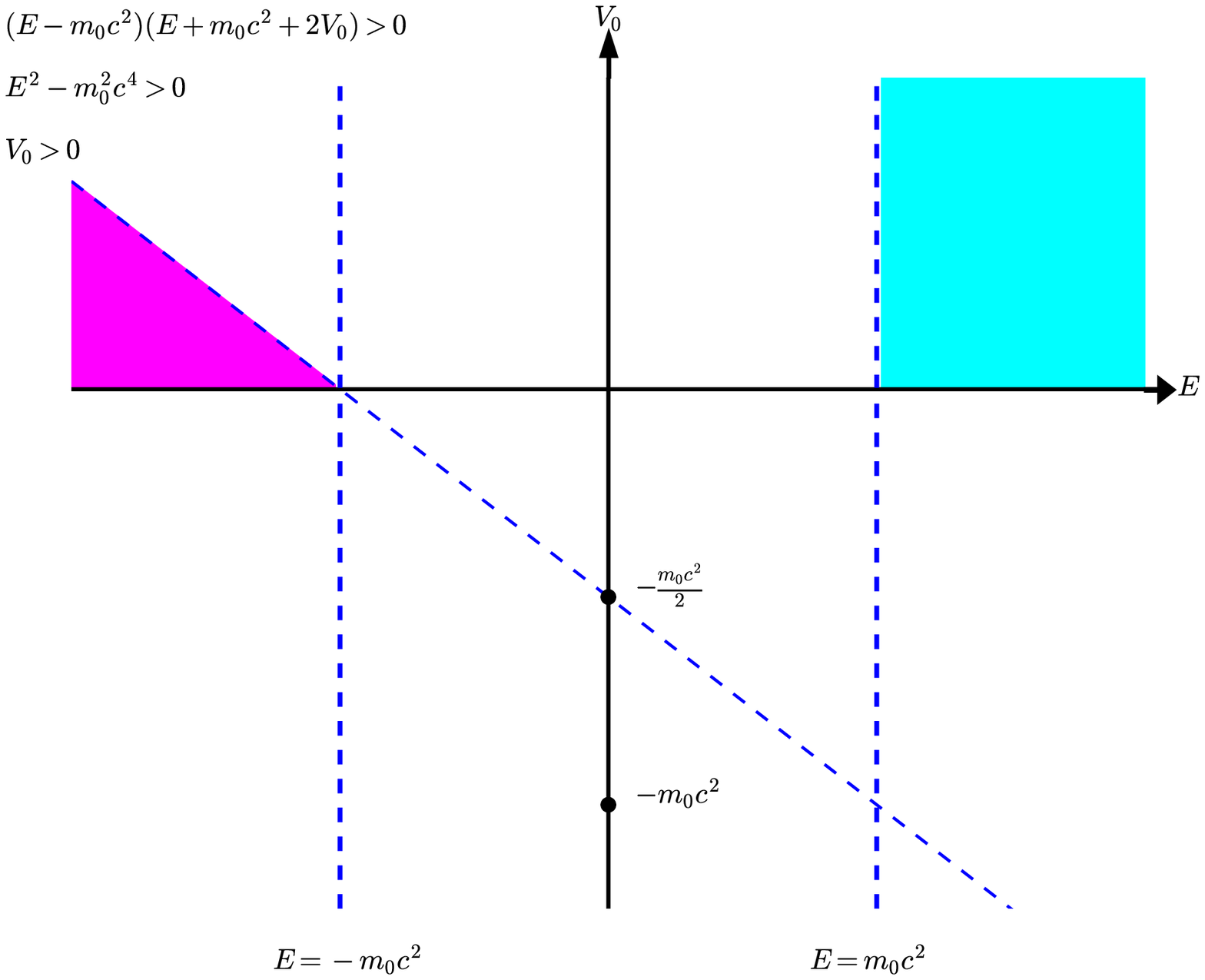}
\caption{Possible energy regions and forbidden energy gap for particle scattering in PSS case.}
\label{fig:2}       
\end{figure*}

Let us now consider the case with the negative sign as plotted in Fig.~\ref{fig:2}. There is no change in the conditions on the energy for the scattering of the particle with positive energy.  This region is indicated by cyan color in Fig.~\ref{fig:2}. When $V_0=0$, the energy \emph{needed for scattering} for incident particles with negative energy is $-m_0 c^2$. Whereas if $V_0$ exceeds zero, these particles that have energies smaller than $-m_0 c^2$ can scatter. For particles with negative energy this region is represented by the magenta area in  Fig.~{\ref{fig:2}}.

\subsection{Asymptotic Behaviors}

To proceed further, we investigate the behavior of the wave function at infinities. In both limits, the wave functions do not depend on hypergeometric functions. Hence the surface terms do not possess an effect on them. This is very reasonable since the incoming and outgoing relativistic particles do not feel any effect while they are far from the target particle.

In the $x=0$ limit, the wave function should be continuous and well-defined. We clearly find out that the wave function depends on the surface effects by $\theta^\mp$ terms within $a^\mp$ and $b^\mp$ inside the hypergeometric functions too.


\subsubsection{$x \rightarrow -\infty$ limit}
In the negative region, at this limit the wave function behaves as
\begin{eqnarray}
 \phi_L^\mp(x\rightarrow -\infty) &\approx& e^{-\frac{\pi \kappa_\mp}{a}} \Big(D_1^\mp e^{ik(x+L)}+D_2^\mp e^{-ik(x+L)} \Big)
\end{eqnarray}

\subsubsection{$x\rightarrow 0^-$ case}

In the negative domain, when the wave function approaches to the origin, the hypergeometric function plays an important role. In this limit
\begin{eqnarray}
  z\Big|_{x\rightarrow 0^-} &=& (1+e^{-aL})^{-1} \equiv t_0\,.
\end{eqnarray}
Since $aL\gg 1$ the new constant $t_0\approx 1$. Therefore we need to use the identity
\begin{eqnarray}\label{HypergeoZ=1}
  _2F_1(a,b,c;t) &=& \frac{\Gamma(c)\Gamma(c-a-b)}{\Gamma(c-a)\Gamma(c-b)} \,\,\, _2F_1(a,b, a+b-c+1;1-t)+(1-t)^{c-a-b}\nonumber \\
  &&\times \frac{\Gamma(c)\Gamma(a+b-c)}{\Gamma(a)\Gamma(b)}\,\,\, _2F_1(c-a,c-b,
  c-a-b+1;1-t),\,\,\,\,\,\,\,\,\,\,\,\,
\end{eqnarray}
for the  transformation of the hypergeometric function \cite{RefGradshytenRyzhikBook}. Then we can simply give the wave function behavior as it approaches to zero as follows
\begin{eqnarray}
 \phi_L^\mp(x\rightarrow 0^-) &=& e^{-\frac{\pi \kappa_\mp}{a}}\Bigg[D_1^\mp t_0^\mu \bigg[(1-t_0)^{\nu_\mp} S_1^\mp N_1^\mp+ (1-t_0)^{-\nu_\mp} S_2^\mp N_2^\mp \bigg] \nonumber \\
 &&+ D_2^\mp t_0^{-\mu} \bigg[(1-t_0)^{\nu_\mp} S_3^\mp N_3^\mp+ (1-t_0)^{-\nu_\mp} S_4^\mp N_4^\mp \bigg] \Bigg].
\end{eqnarray}
where
\begin{eqnarray}
  S_1^\mp &\equiv& \frac{\Gamma(1+2\mu)\Gamma(-2\nu_\mp)}{\Gamma(1+\mu-\theta_\mp-\nu_\mp)\Gamma(\mu+\theta_\mp-\nu_\mp)}\,,\\
  S_2^\mp &\equiv& \frac{\Gamma(1+2\mu)\Gamma(2\nu_\mp)}{\Gamma(1+\mu-\theta_\mp+\nu_\mp)\Gamma(\mu+\theta_\mp+\nu_\mp)} \,,\\
  S_3^\mp &\equiv& \frac{\Gamma(1-2\mu)\Gamma(-2\nu_\mp)}{\Gamma(1-\mu-\theta_\mp-\nu_\mp)\Gamma(-\mu+\theta_\mp-\nu_\mp)}\,, \\
  S_4^\mp &\equiv& \frac{\Gamma(1-2\mu)\Gamma(2\nu_\mp)}{\Gamma(1-\mu-\theta_\mp+\nu_\mp)\Gamma(-\mu+\theta_\mp+\nu_\mp)}\,,
\end{eqnarray}
and
\begin{eqnarray}
  N_1^\mp &\equiv&  \,\,\, _2F_1[\mu+\theta_\mp+\nu_\mp,1+\mu-\theta_\mp+\nu_\mp, 1+2\nu_\mp;1-t_0]\,, \\
  N_2^\mp &\equiv&  \,\,\, _2F_1[1+\mu-\theta_\mp-\nu_\mp,\mu+\theta_\mp-\nu_\mp, 1-2\nu_\mp;1-t_0]\,, \\
  N_3^\mp &\equiv&  \,\,\, _2F_1[-\mu+\theta_\mp+\nu_\mp,1-\mu-\theta_\mp+\nu_\mp,1+2\nu_\mp;1-t_0]\,, \\
  N_4^\mp &\equiv&  \,\,\, _2F_1[1-\mu-\theta_\mp-\nu_\mp,-\mu+\theta_\mp-\nu_\mp, 1-2\nu_\mp;1-t_0]\,.
\end{eqnarray}
The wave function behaves near zero as
\begin{eqnarray}
  \phi_L^\mp(x\rightarrow 0^-) &\approx& e^{-\frac{\pi \kappa_\mp}{a}}\Bigg[\Big(D_1^\mp S_1^\mp N_1^\mp+D_2^\mp S_3^\mp N_3^\mp\Big)e^{-i\kappa_\mp (x+L)} \nonumber \\
  &&+ \Big(D_1^\mp S_2^\mp N_2^\mp+D_2^\mp S_4^\mp N_4^\mp \Big) e^{i\kappa_\mp (x+L)} \Bigg].  \,\,\,\,\,\,\,\,\,\,\,\,
\end{eqnarray}

\subsubsection{$x\rightarrow 0^+$ case}
In the positive region, the wave function traveling from the origin to the right through the potential well shows a similar behavior as on the other side of the potential well, that is mentioned above. Thus,
\begin{eqnarray}
  \phi_R^\mp(x\rightarrow 0^+) &\approx& e^{-\frac{\pi \kappa_\mp}{a}} \Bigg[\Big(D_3^\mp S_1^\mp N_1^\mp+D_4^\mp S_3^\mp N_3^\mp\Big)e^{i\kappa_\mp (x-L)} \nonumber \\
  &&+ \Big(D_3^\mp S_2^\mp N_2^\mp+D_4^\mp S_4^\mp N_4^\mp \Big) e^{-i\kappa_\mp (x-L)} \Bigg].  \,\,\,\,\,\,\,\,\,\,\,\,
\end{eqnarray}

\subsubsection{$x \rightarrow \infty$ limit}
The asymptotic behavior in the positive infinity is found as
\begin{eqnarray}
  \phi_R^\mp(x\rightarrow \infty) & \approx & e^{-\frac{\pi \kappa_\mp}{a}} \Big[D_3^\mp e^{-ik(x-L)} +D_4^\mp e^{ik(x-L)} \Big].
\end{eqnarray}

Since the physics is independent of the choice of incident particle direction, in this study we choose the particle to approach the target from the negative region. Therefore in this region, we may have only the transmitted wave, so $D_3^\mp$ is equal to zero.
\subsection{The Continuity Conditions }
The wave functions must have smooth behavior at every point in the configuration space. In our problem, this conditions should be satisfied by
\begin{eqnarray}
  \phi_L^\mp(x) \bigg|_{x=0^-} &=& \phi_R^\mp(x) \bigg|_{x=0^+}, \label{kendisi} \\
  \frac{d \phi_L^\mp(x)}{dx} \bigg|_{x=0^-} &=& \frac{d \phi_R^\mp(x)}{dx} \bigg|_{x=0^+}. \label{turevi}
\end{eqnarray}
By employing Eq.~(\ref{kendisi}) we find
\begin{eqnarray}
  \frac{D_4^\mp}{D_1^\mp}- \frac{D_2^\mp}{D_1^\mp} &=& t_0^{2\mu} \frac{M_1^\mp}{M_2^\mp}\,,   \label{ilk kosul}
\end{eqnarray}
where
\begin{eqnarray}
  M_1^\mp &\equiv&   S_1^\mp N_1^\mp+ (1-t_0)^{-2\nu_\mp} S_2^\mp N_2^\mp, \\
  M_2^\mp &\equiv&   S_3^\mp N_3^\mp+ (1-t_0)^{-2\nu_\mp} S_4^\mp N_4^\mp.
\end{eqnarray}
Instead of the second condition given by Eq.~(\ref{turevi}), we can use
\begin{eqnarray}
   \frac{d \phi_L^\mp(z)}{dz} \bigg|_{z=t_0} &=& -\frac{d \phi_R^\mp(y)}{dy} \bigg|_{y=t_0}. \label{tureviyeni}
\end{eqnarray}
and after straightforward calculations we get
\begin{eqnarray}
  \frac{D_4^\mp}{D_1^\mp}+ \frac{D_2^\mp}{D_1^\mp} &=& - t_0^{2\mu} \frac{\Bigg[\bigg( \frac{\mu}{t_0}+\frac{\nu_\mp}{t_0-1}\bigg)M_1^\mp+\frac{(\mu+\theta_\mp+\nu_\mp)(1+\mu-\theta_\mp+\nu_\mp)}{1+2\mu}M_3^\mp\Bigg]}{\Bigg[\bigg( \frac{-\mu}{t_0}+\frac{\nu_\mp}{t_0-1}\bigg)M_2^\mp+\frac{(-\mu+\theta_\mp+\nu_\mp)(1-\mu-\theta_\mp+\nu_\mp)}{1-2\mu}M_4^\mp\Bigg]}. \,\,\,\,
  \label{ikinci kosul}
\end{eqnarray}
where
\begin{eqnarray}
  M_3^\mp &\equiv&   S_5^\mp N_5^\mp+ (1-t_0)^{-1-2\nu_\mp} S_6^\mp N_6^\mp, \\
  M_4^\mp &\equiv&   S_7^\mp N_7^\mp+ (1-t_0)^{-1-2\nu_\mp} S_8^\mp N_8^\mp.
\end{eqnarray}
Here
\begin{eqnarray}
  S_5^\mp &\equiv& \frac{\Gamma(2+2\mu)\Gamma(-1-2\nu_\mp)}{\Gamma(1+\mu-\theta_\mp-\nu_\mp)\Gamma(\mu+\theta_\mp-\nu_\mp)} \,,\\
  S_6^\mp &\equiv& \frac{\Gamma(2+2\mu)\Gamma(1+2\nu_\mp)}{\Gamma(1+\mu+\theta_\mp+\nu_\mp)\Gamma(2+\mu-\theta_\mp+\nu_\mp)}\,, \\
  S_7^\mp &\equiv& \frac{\Gamma(2-2\mu)\Gamma(-1-2\nu_\mp)}{\Gamma(1-\mu-\theta_\mp-\nu_\mp)\Gamma(-\mu+\theta_\mp-\nu_\mp)}\,, \\
  S_8^\mp &\equiv& \frac{\Gamma(2-2\mu)\Gamma(1+2\nu_\mp)}{\Gamma(1-\mu+\theta_\mp+\nu_\mp)\Gamma(2-\mu-\theta_\mp+\nu_\mp)}
\end{eqnarray}
and
\begin{eqnarray}
  N_5^\mp &\equiv&  \,\,\, _2F_1[1+\mu+\theta_\mp+\nu_\mp,2+\mu-\theta_\mp+\nu_\mp, 2+2\nu_\mp;1-t_0]\,, \\
  N_6^\mp &\equiv&  \,\,\, _2F_1[1+\mu-\theta_\mp-\nu_\mp,\mu+\theta_\mp-\nu_\mp, -2\nu_\mp;1-t_0] \,,\\
  N_7^\mp &\equiv&  \,\,\, _2F_1[1-\mu+\theta_\mp+\nu_\mp,2-\mu-\theta_\mp+\nu_\mp,2+2\nu_\mp;1-t_0] \,,\\
  N_8^\mp &\equiv&  \,\,\, _2F_1[1-\mu-\theta_\mp-\nu_\mp,-\mu+\theta_\mp-\nu_\mp, -2\nu_\mp;1-t_0]\,.
\end{eqnarray}
Before proceeding further, we would like to stress that the equality
\begin{eqnarray}
  \frac{d}{dt}\,\,\, _2F_1[a,b,c,t]&=& \frac{ab}{c}  \,\,\, _2F_1[a+1,b+1,c+1,t]
\end{eqnarray}
is needed to be used and in some papers it has been ignored \cite{RefPanella2010, LutfuogluAkdeniz2016}. In those studies, the absence of this term can be noticed when carefully looking at the plot of the wave-functions.

Finally, we would like to emphasize that the continuity conditions that are derived in Eq.~(\ref{ilk kosul}) and Eq.~(\ref{ikinci kosul}) can be used to find the resonant states with the Siegert boundary conditions \cite{RefHatano2009}.

\subsection{The Reflection and Transmission Coefficients}

The reflection coefficient $R^\mp$,
\begin{eqnarray}
  R^\mp &=& \frac{D_2^\mp}{D_1^\mp},
\end{eqnarray}
and the transmission coefficient $T^\mp$,
 \begin{eqnarray}
  T^\mp &=& \frac{D_4^\mp}{D_1^\mp},
\end{eqnarray}
can be obtained by solving Eq.~(\ref{ilk kosul}) and Eq.~(\ref{ikinci kosul}).
\begin{eqnarray}
  R^\mp &=& -\frac{t_0^{2\mu}}{2}\Bigg[ \frac{\Big( \frac{\mu}{t_0}+\frac{\nu_\mp}{t_0-1}\Big)M_1^\mp+\frac{(\mu+\theta_\mp+\nu_\mp)(1+\mu-\theta_\mp+\nu_\mp)}{1+2\mu}M_3^\mp}{\Big( \frac{-\mu}{t_0}+\frac{\nu_\mp}{t_0-1}\Big)M_2^\mp+\frac{(-\mu+\theta_\mp+\nu_\mp)(1-\mu-\theta_\mp+\nu_\mp)}{1-2\mu}M_4^\mp}+\frac{M_1^\mp}{M_2^\mp}\Bigg]. \,\,\,\, \label{refcoef}\\
  T^\mp &=& -\frac{t_0^{2\mu}}{2}\Bigg[ \frac{\Big( \frac{\mu}{t_0}+\frac{\nu_\mp}{t_0-1}\Big)M_1^\mp+\frac{(\mu+\theta_\mp+\nu_\mp)(1+\mu-\theta_\mp+\nu_\mp)}{1+2\mu}M_3^\mp}{\Big( \frac{-\mu}{t_0}+\frac{\nu_\mp}{t_0-1}\Big)M_2^\mp+\frac{(-\mu+\theta_\mp+\nu_\mp)(1-\mu-\theta_\mp+\nu_\mp)}{1-2\mu}M_4^\mp}-\frac{M_1^\mp}{M_2^\mp}\Bigg]. \,\,\,\, \label{transcoef}
\end{eqnarray}
The physically meaningful quantities, which express the probabilities that the incident particles are reflected or transmitted from the potential barrier, are $\Big|R^\mp\Big|^2$ and $\Big|T^\mp\Big|^2$. One can prove by the standard techniques that for $|E|\geq m_0 c^2$ the total probability is equal to unity since we have $k \in \mathbb{R}$.


\section{Results}\label{sec:res}
As an example, in the strong regime, we consider a neutral kaon scattering from a nucleus whose the reciprocal diffusion parameter and the nuclear radius equal to $1 \,(\mathrm{fm}^{-1})$ and $6\,(\mathrm{fm})$, respectively. Note these assignments satisfy our beginning assumption, $aL \gg 1$.
\subsection{Spin Symmetric Case}
The graphical analysis of the SS case given in Fig.~\ref{fig:1} dictates us to investigate two different cases. Therefore, we choose the parameter $V_0$ to be smaller or greater than the rest mass energy of the neutral kaon.

In Fig.~\ref{SSC_V0_0,5mc^2-W0_0,0mc^2}, Fig.~\ref{SSC_V0_0,5mc^2-W0_0,5mc^2},  Fig.~\ref{SSC_V0_0,5mc^2-W0_2,0mc^2}, and Fig.~\ref{SSC_V0_0,5mc^2-W0_4,0mc^2} the reflection and transmission probabilities versus incident particle energies are given. In these graphs, the first row is plotted to show the forbidden gap where the non-physical solutions are found. In the second row, to analyze the physical scattering processes better, the first and the second column are studied separately for the scattering of particles with positive and negative energy.

In the case $V_0=\frac{m_0 c^2}{2}$, in the absence of surface effects $W=0$ or when these effects are not strong enough to build a barrier $W=\frac{m_0 c^2}{2}$, the particles with positive and negative energy can be scattered except the forbidden KG gap shown in Fig.~\ref{SSC_V0_0,5mc^2-W0_0,0mc^2} and Fig.~\ref{SSC_V0_0,5mc^2-W0_0,5mc^2}. When the surface effects increase, particles need momentum to scatter or tunnel from the barrier. Therefore, the absolute value of the relativistic energy should increase. For the particle with negative energy, such increase of the energy does not affect the probability of scattering as shown in the second row first column in Fig.~\ref{SSC_V0_0,5mc^2-W0_2,0mc^2} and Fig.~\ref{SSC_V0_0,5mc^2-W0_4,0mc^2}.
This is not valid for the scattering of particles with positive energy. There is an absolute shift of the scatterable particle energies. Moreover, due to the quantum effects there arise resonances as shown in the second row and column of the  Fig.~\ref{SSC_V0_0,5mc^2-W0_2,0mc^2} and Fig.~\ref{SSC_V0_0,5mc^2-W0_4,0mc^2}.

In the case $V_0=\frac{3m_0 c^2}{2}$, the KG forbidden gap expands through the negative energy side. The scatterable particle energies shift to the energies smaller than $-2m_0c^2$ as predicted in Fig.~\ref{fig:1}. Even if there is no surface effect, $W=0$, or no barrier yet, $W=\frac{3m_0 c^2}{2}$, the transmission and reflection probability of particles with negative energy are given in the second row in the first column in Fig.~\ref{SSC_V0_1,5mc^2-W0_0,0mc^2} and Fig.~\ref{SSC_V0_1,5mc^2-W0_1,5mc^2}. The transmission probability of the particle with positive energy starts to change slightly even there is no barrier yet. The main reason of this is the change of the shape of the potential well, more precisely the squeezing of the well. When surface effects start to create the barrier, particles need extra energy for transmission. Therefore, the energies of transitable particles increase as shown in second row second column of Fig.~\ref{SSC_V0_1,5mc^2-W0_2,5mc^2} and  Fig.~\ref{SSC_V0_1,5mc^2-W0_4,0mc^2}. For deeper well, the quantum effects, in other words, the resonance energy states, are found with smaller energy values.

\subsection{Pseudo-spin Symmetric Case}

The PSS differs from the SS in means of the shifting of the rest mass energy,
\begin{eqnarray}
   m_0^{\mp} &\rightarrow & m_0 \mp \frac{V_v}{c^2}.
\end{eqnarray}
Therefore in SS or PSS cases particles have different effective rest masses. More precisely, in the strong regime, in SS case effective mass increases while in PSS case decreases.

On the other hand, in Fig.~\ref{fig:2} we have shown in scattering case that with a positive $V_0$ parameter the forbidden KG gap is widening through the negative energy values. To be able to compare it with the SS case, we take the $V_0$ parameter as half of the rest energy of neutral kaon.  In the absence of surface effects, the KG forbidden gap is shown in Fig.~\ref{PSSC_V0_0,5mc^2-W0_0,0mc^2}. The scattering probability of the particle with negative energy is found to be unity after the shift as predicted by Fig.~\ref{fig:2}.

When the surface effects are taken into account, but yet cannot build a barrier, there are slight changes in the particle energy and they can transmit as shown in Fig.~\ref{PSSC_V0_0,5mc^2-W0_0,5mc^2}.

If the surface interactions are more effective, we find out that negative-energy side does not vary as in SS case. On the other hand, the transmission probabilities of the positive-energy particles start to fluctuate just after $m_0c^2$ as shown in  Fig.~\ref{PSSC_V0_0,5mc^2-W0_2,0mc^2} and Fig.~\ref{PSSC_V0_0,5mc^2-W0_4,0mc^2}.

\section{Conclusion}\label{sec:conclusions}
We found the transmission and reflection probabilities for a Klein-Gordon (KG) particle in both Spin Symmetric (SS) and Pseudo-Spin Symmetric (PSS) case. In the SS case, the increase of the potential $W$ (the surface effect) does not considerably influence the threshold for scattering for the particles with negative energy; the threshold is given by the potential $V_0$. On the other hand, for positive energies it holds true that the bigger the surface effect is, the bigger the energy needed for scattering must be. In the PSS case the rest mass energy is shifted. Again, the transmission and reflection probabilities are affected by the surface effect mainly in the positive-energy region. For both SS and PSS case one can observe resonances for positive energies, which are more pronounced for bigger $W$.

\begin{acknowledgements}
This work was partially supported by the Turkish Science and Research Council (TUBITAK) and Akdeniz University. J. L. was supported by the grant 15-14180Y of the Czech Science Foundation. Authors would like to thank Can Ertugay for his help for the preparation of the plots  $Fig.~\ref{fig:1}$ and $Fig.~\ref{fig:2}$ and Dr. Esat Pehlivan for proofreading. Finally, we would like to emphasize our gratitude to the anonymous referee for his/her careful review of our manuscript with very kind comments, corrections and suggestions.
\end{acknowledgements}


\begin{thebibliography}{}
%
%

\bibitem{RefBohr1982} A. Bohr, I. Hamamoto, B. R. Mottelson, Phys. Scr., 26, 267--272 (1982).
\bibitem{RefMottelson1991}  B. Mottelson, Nucl. Phys. , A522, 1c--12c (1991).
\bibitem{RefZhouMengRing2003} S.Zhou, J. Meng, P. Ring, Phys. Rev. Lett, 91, 262501, (2003).
%

%
\bibitem{RefSmithTassie1971} G. B. Smith, L. J. Tassie, Ann. Phys., 65, 352--360 (1971).
\bibitem{RefBellRuegg1975} J. S. Bell, H. Ruegg, Nucl. Phys., B98, 151--153 (1975).
%

\bibitem{RefGinocchio1997} J. N. Ginocchio, Phys. Rev. Lett., 78, 436--439 (1997).
\bibitem{RefGinocchio1999} J. N. Ginocchio, Phys. Rep., 315, 231--240 (1999).
\bibitem{RefCastro2005} A. S. de Castro, Phys. Lett. A, 338, 81--89 (2005).
\bibitem{RefAlhaidariBahlouliAl-Hasan2006}
A. D. Alhaidari, H. Bahloul, A. Al-Hasan, Phys. Lett. A, 349, 87--97 (2006).
\bibitem{RefAlbertoCastroMalheiro2007} P. Alberto, A. S. de Castro, M. Malheiro, Phys. Rev. C, 75, 047303 (2007).
\bibitem{RefAlbertoCastroMalheiro2013} P. Alberto, A. S. de Castro, M. Malheiro, Phys. Rev. C, 87, 031301(R) (2013).
\bibitem{RefHamzaviRijabi2013}M. Hamzavi and A. A. Rajabi, ISRN High Energy Physics, Vol. 2013, 987632 (2013)
\bibitem{RefAlbertoMalheiroFredericoCastro2015}
P. Alberto, M. Malheiro, T. Frederico, A. S. de Castro, Phys. Rev. A, 92, 062137 (2015).
\bibitem{RefIkotetal2015}
A. N. Ikot, H. Hassanabadi, T. M. Abbey, Commun. Theor. Phys., 64, 637--643, (2015).
\bibitem{RefAlbertoetal2016}
P. Alberto, M. Malheiro, T. Frederico, A. de Castro, J. Phys.: Conf. Ser., 738, 012033, (2016).
%
%
\bibitem{RefGinocchio2005}
J. N. Ginocchio, Phys. Rep., 414, 165--261 (2005).
\bibitem{RefLiangMengZhou2015}
H. Liang, J. Meng, S. Zhou, Phys. Rep., 570, 1--84 (2015).
\bibitem{RefBlokhinBahriDraayer1995}
A. L. Blokhin, C. Bahri, J. P. Draayer, Phys. Rev. Lett., 74, 4149--4152 (1995).
\bibitem{RefHamzavi2010}
M. Hamzavi, A.A. Rajabi, H. Hassanabadi, Phys. Lett., A374, 4303--4307 (2010).
\bibitem{RefHassanabadiIkotYarrinkamar2014}
H. Hassanabadi, A.N. Ikot, S. Zarrinkamar, Acta Phys. Pol. A, 126 (2014).

\bibitem{RefIkotAtAll2015}
A.N. Ikot, H.P. Obong, I.O. Owate, M.C. Onyeaju, H. Hassanabadi, Adv. High Energy Phys.,  2015, 632603 (2015)
\bibitem{RefHartmann2017} R.R. Hartmann, M.E. Portnoi, Sci. Rep., 7, 11599 (2017).
\bibitem{RefIkotAtAll2016}
A.N. Ikot, H.P. Obong, T.M. Abbey, S. Zare, M. Ghafourian, H. Hassanabadi, Few-Body Syst 57, 807--822 (2016)
\bibitem{RefGaoZhang2017}
J. Gao, M.C. Zhang, Phys. Lett., B769, 77--81 (2017).
\bibitem{RefXieJia2015}
X.J. Xie, C.S. Jia, Phys. Scr., 90, 035207 (2015)
\bibitem{RefZhanetal2016}
P. Zhang, H.C. Long, C.S. Jia, Eur. Phys. J. Plus., 131, 117 (2016)


\bibitem{RefKennedy2002}
P. Kennedy, J. Phys., A35, 689--698 (2002).
\bibitem{RefPanella2010}
O. Panella, S. Biondini, A. Arda, J. Phys. A: Math. Theor., 43, 325302 (2010).
\bibitem{RefOlgarMutaf2015}
E. Olgar, H. Mutaf, Adv. Math. Phys., 2015, 923076 (2015).

\bibitem{RefRojasVIllalba}
C. Rojas, V.M. Villalba, Phys. Rev. A 71, 052101 (2005)
\bibitem{HassanabadiMaghsoodiZarrinkamarSalehi2013}
H. Hassanabadi, E. Maghsoodi, S. Zarrinkamar, N. Salehi, Few-Body Sys., 54, 2009--2012 (2013).

\bibitem{RefWoodsSaxon1954}
R. D. Woods, D. S. Saxon, Phys. Rev., 95, 577--578 (1954)
\bibitem{RefTianWangLi2007}
J.-L. Tian, N. Wang, Z.-X. Li, Chin. Phys. Lett., 24, 905--908 (2007)
\bibitem{CandemirBayrak2014} N. Candemir, O. Bayrak, Mod. Phys. Lett. A., {\bf29}: 1450180 (2014).
\bibitem{BayrakAciksoz2015} O. Bayrak, E. Aciksoz, Phys. Scr., {\bf90}: 015302 (2015).
\bibitem{BayrakSahin2015} O. Bayrak, D. Sahin,Commun. Theor. Phys., {\bf64}: 259--262 (2015).
\bibitem{LutfuogluAkdeniz2016} B. C. L\"{u}tf\"{u}o\u{g}lu, F. Akdeniz, O. Bayrak, J. Math. Phys., {\bf57}, 032103 (2016).
\bibitem{RefLutfuoglu2018} B. C. L\"{u}tf\"{u}o\u{g}lu, Commun. Theor. Phys., {\bf69}: 23--27  (2018).

\bibitem{LiendoCastro2016} J. A. Liendo, E. Castro, R. Gomez et al, Int. J. Mod. Phys. E, {\bf225}: 1650055 (2016).
\bibitem{BerkdemirBerkdemir2005} C. Berkdemir, A. Berkdemir, R. Sever,  Phys. Rev. C, {\bf72}: 027001 (2005),  Errata {\bf74}: 027001 (2005).
\bibitem{BadalovAhmado2009} V. H. Badalov, H. I. Ahmadov, A. I. Ahmadov, Int. J. Mod. Phys. E, {\bf18}: 631 (2009).
\bibitem{GonulKoksal2007} B. G\"{o}n\"{u}l, K. K\"{o}ksal, Phys. Scr., {\bf76}: 565--570 (2007).
\bibitem{KouraYamada2000} H. Koura, M. Yamada, Nucl. Phys. A,  {\bf671}: 96--118 (2000).
\bibitem{CapakPetrellis2015}  M. \c{C}apak, D. Petrellis, B. G\"{o}n\"{u}l et al, J. Phys. G, {\bf42}: 95102(2015).
\bibitem{CapakGonul2016} M. \c{C}apak, G\"{o}n\"{u}l, Mod. Phys. Lett. A, {\bf31}: 1650134 (2016).
\bibitem{IkotAkpan2012} A. N. Ikot, I. O. Akpan, Chin. Phys. Lett., {\bf29}: 090302 (2012).
\bibitem{IkhdairFalayeHamzavi2013} S. M. Ikhdair, B. J. Falaye, M. Hamzavi, Chin. Phys. Lett., {\bf30}: 020305 (2013).
\bibitem{surface1} A. M. Kobos, R. S. Mackintosh, Phys. Rev. C, {\bf26}: 1766--1769 (1982).
\bibitem{surface2} I. Boztosun, Phys. Rev. C., {\bf66}: 024610 (2002).
\bibitem{surface3} I. Boztosun, O. Bayrak, Y. Dagdemir,  Int. J. Mod. Phys. E, {\bf14}: 663--673 (2005).
\bibitem{surface4} G. Kocak, M. Karakoc, I. Boztosun et al, Phys. Rev. C, {\bf81}: 024615 (2010).
\bibitem{surface5} H. Dapo, I. Boztosun, G. Kocak et al,  Phys. Rev. C, {\bf85}: 044602 (2012).
\bibitem{RefKandirmaz2017}
N. Kandirmaz, GU J Sci 30, 133--138 (2017)
\bibitem{RefGradshytenRyzhikBook}
I.S. Gradshteyn, I.M. Ryzhik, Table of Integrals, Series and Products, 7th Ed., Academic Press, Elsevier, USA.
\bibitem{RefHatano2009}
N. Hatano, T. Kawamoto, J. Feinberg, J. Pramana - J Phys 73, 553 (2009).

\end{thebibliography}

\newpage


\newpage

\newpage

\newpage
\begin{figure*}
  \includegraphics[width=1.0\textwidth]{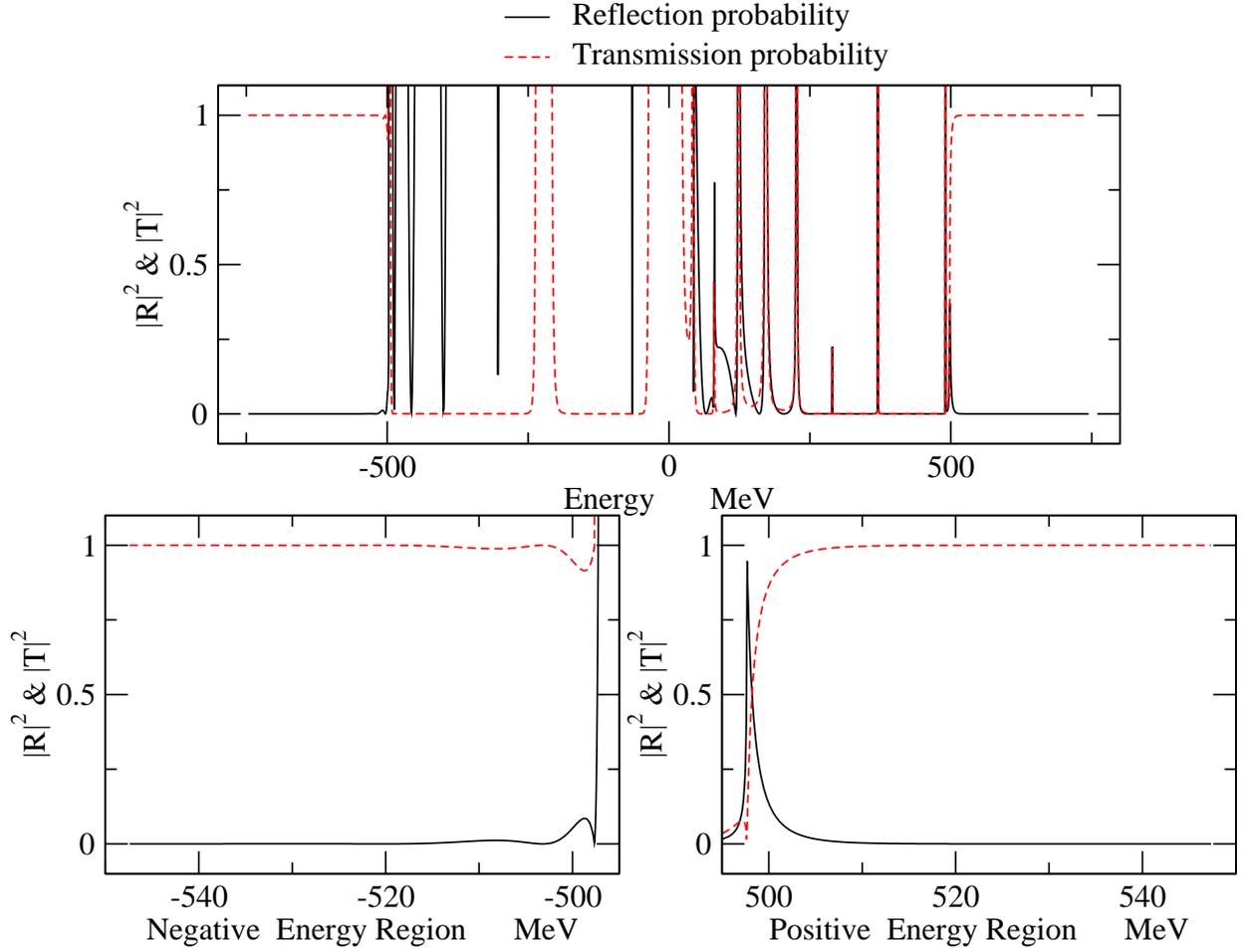}
\caption{In case of SS, the plots of the reflection $R$ and transmission $T$ probabilities versus incident particle energies  which has rest mass energy $m_0c^2=497.648\,\mathrm{MeV}$. The red dashed lines are the transmission, while the black solid ones are the reflection probabilities.  The potential parameters are chosen arbitrarily as $V_0=\frac{m_0c^2}{2}$, $W=0$, $a=1\,\mathrm{fm}^{-1}$,  $L=6\,\mathrm{fm}$. In the first line we see the KG forbidden gap. In the second line, the particle scattering graphs are given in more details. }
\label{SSC_V0_0,5mc^2-W0_0,0mc^2}       
\end{figure*}

\newpage
\begin{figure*}
  \includegraphics[width=1.0\textwidth]{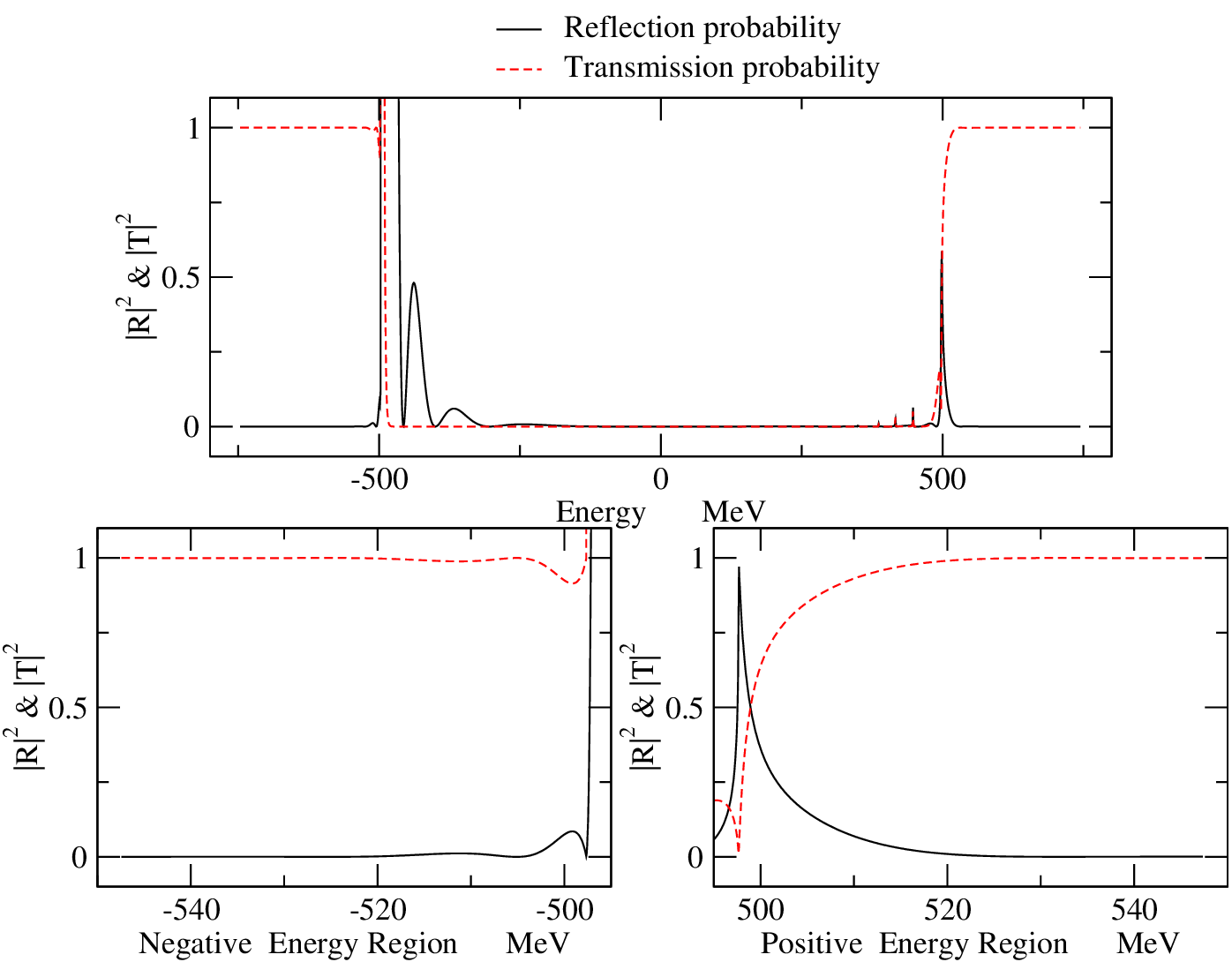}
\caption{In case of SS, the plots of the reflection $R$ and transmission $T$ probabilities versus incident particle energies  which has rest mass energy $m_0c^2=497.648\,\mathrm{MeV}$. The red dashed lines are the transmission, while the black solid ones are the reflection probabilities.  The potential parameters are chosen arbitrarily as $V_0=\frac{m_0c^2}{2}$, $W=\frac{m_0c^2}{2}$, $a=1\,\mathrm{fm}^{-1}$,  $L=6\,\mathrm{fm}$. In the first line we see the KG forbidden gap. In the second line, the particle scattering graphs are given in more details. }
\label{SSC_V0_0,5mc^2-W0_0,5mc^2}       
\end{figure*}

\newpage
\begin{figure*}
  \includegraphics[width=1.0\textwidth]{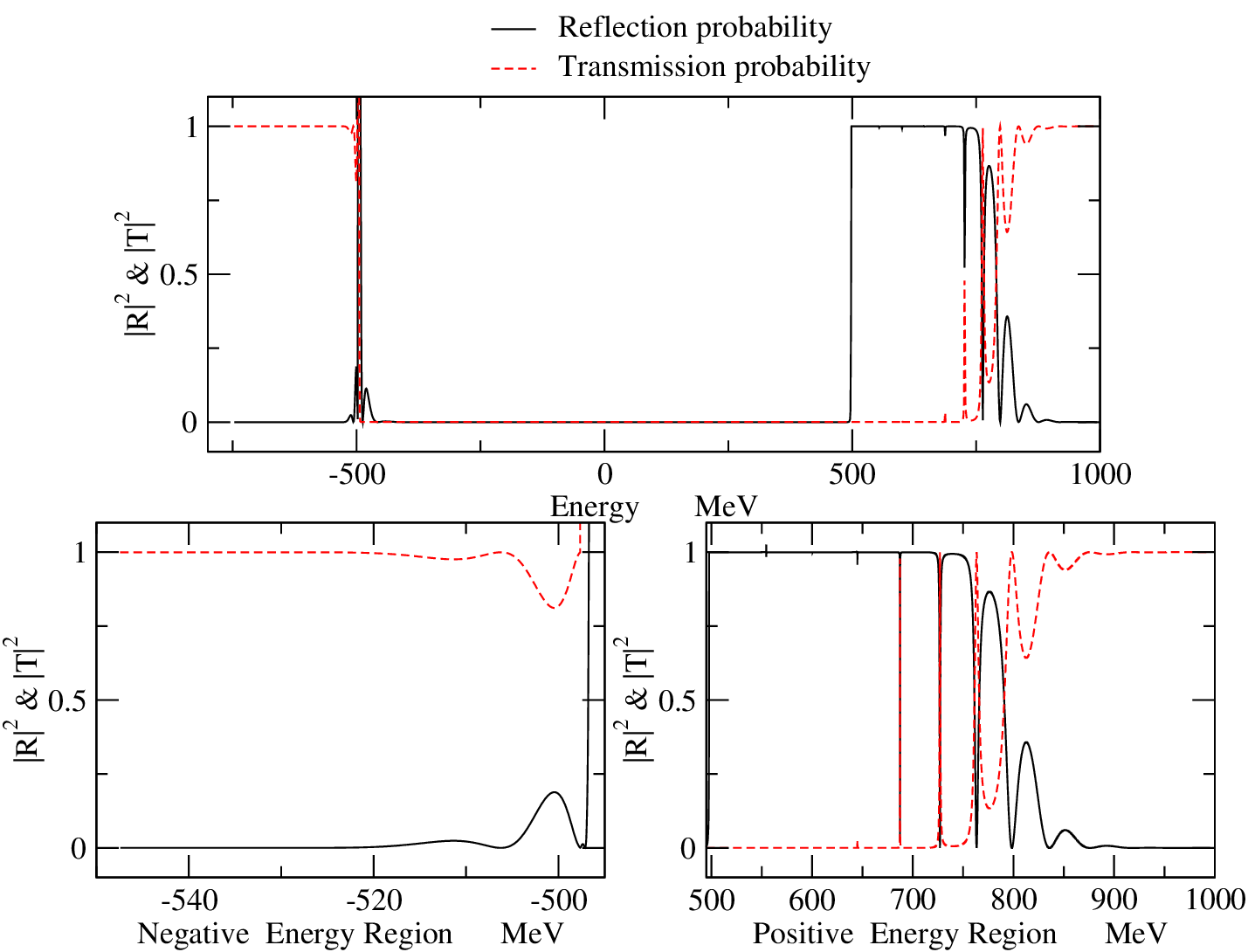}
\caption{In case of SS, the plots of the reflection $R$ and transmission $T$ probabilities versus incident particle energies  which has rest mass energy $m_0c^2=497.648\,\mathrm{MeV}$. The red dashed lines are the transmission, while the black solid ones are the reflection probabilities.  The potential parameters are chosen arbitrarily as $V_0=\frac{m_0c^2}{2}$, $W= 2m_0c^2$, $a=1\,\mathrm{fm}^{-1}$,  $L=6\,\mathrm{fm}$. In the first line we see the KG forbidden gap. In the second line, the particle scattering graphs are given in more details. }
\label{SSC_V0_0,5mc^2-W0_2,0mc^2}       
\end{figure*}

\newpage
\begin{figure*}
  \includegraphics[width=1.0\textwidth]{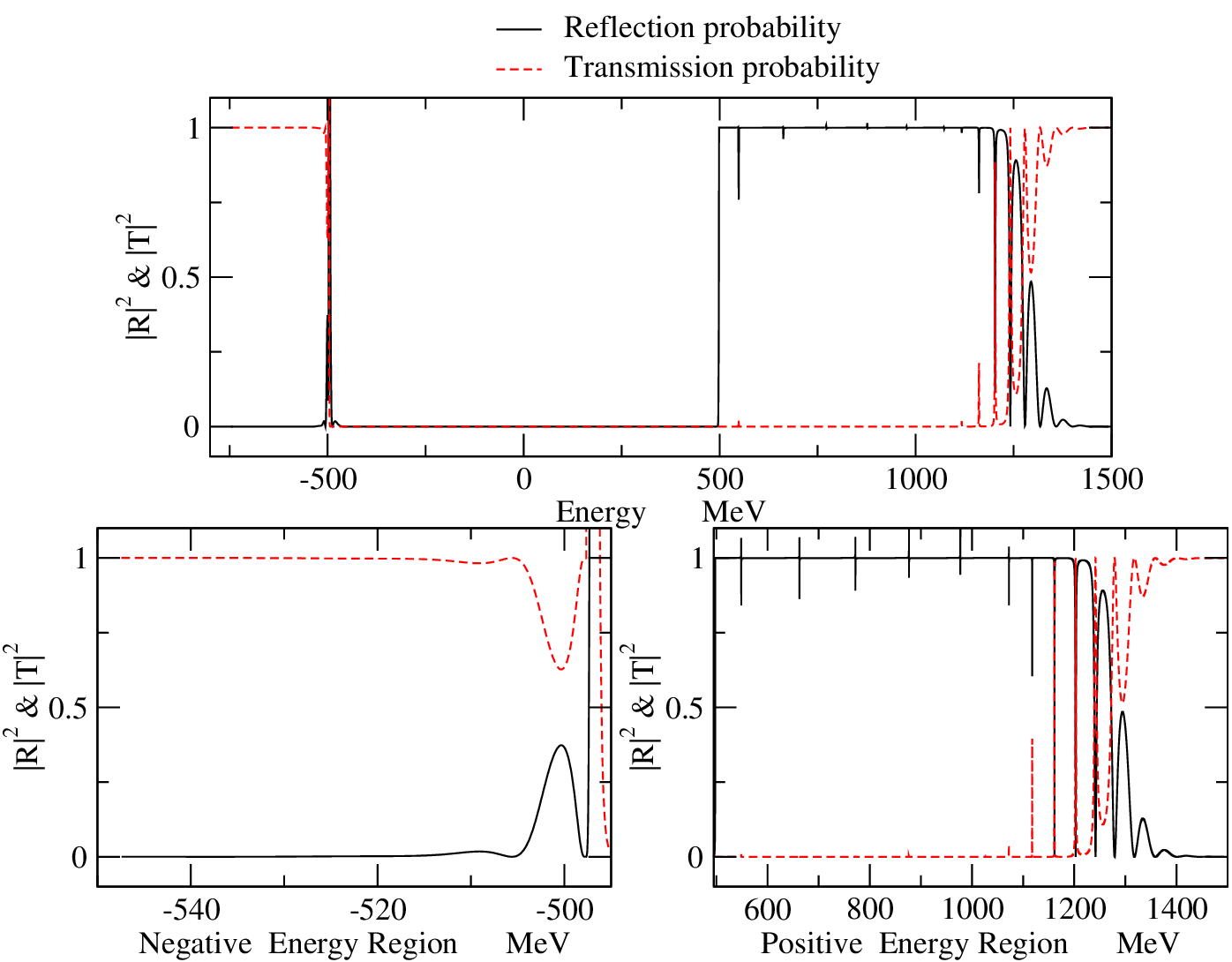}
\caption{In case of SS, the plots of the reflection $R$ and transmission $T$ probabilities versus incident particle energies  which has rest mass energy $m_0c^2=497.648\,\mathrm{MeV}$. The red dashed lines are the transmission, while the black solid ones are the reflection probabilities.  The potential parameters are chosen arbitrarily as $V_0=\frac{m_0c^2}{2}$, $W= 4m_0c^2$, $a=1\,\mathrm{fm}^{-1}$,  $L=6\,\mathrm{fm}$. In the first line we see the KG forbidden gap. In the second line, the particle scattering graphs are given in more details. }
\label{SSC_V0_0,5mc^2-W0_4,0mc^2}       
\end{figure*}

\newpage
\begin{figure*}
  \includegraphics[width=1.0\textwidth]{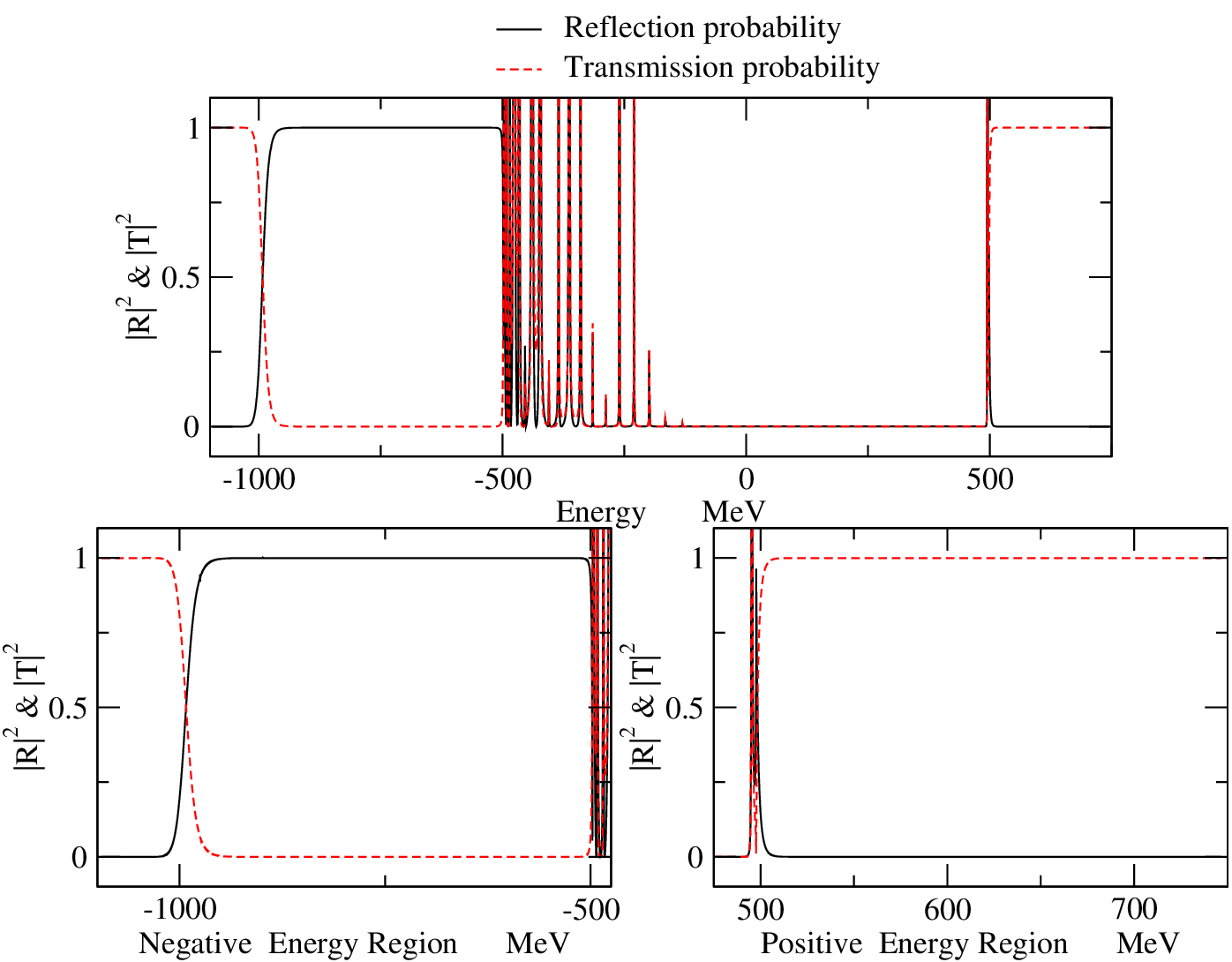}
\caption{In case of SS, the plots of the reflection $R$ and transmission $T$ probabilities versus incident particle energies  which has rest mass energy $m_0c^2=497.648\,\mathrm{MeV}$. The red dashed lines are the transmission, while the black solid ones are the reflection probabilities.  The potential parameters are chosen arbitrarily as $V_0=\frac{3m_0c^2}{2}$, $W=0$, $a=1\,\mathrm{fm}^{-1}$,  $L=6\,\mathrm{fm}$. In the first line we see the KG forbidden gap. In the second line, the particle scattering graphs are given in more details. }
\label{SSC_V0_1,5mc^2-W0_0,0mc^2}       
\end{figure*}

\newpage
\begin{figure*}
  \includegraphics[width=1.0\textwidth]{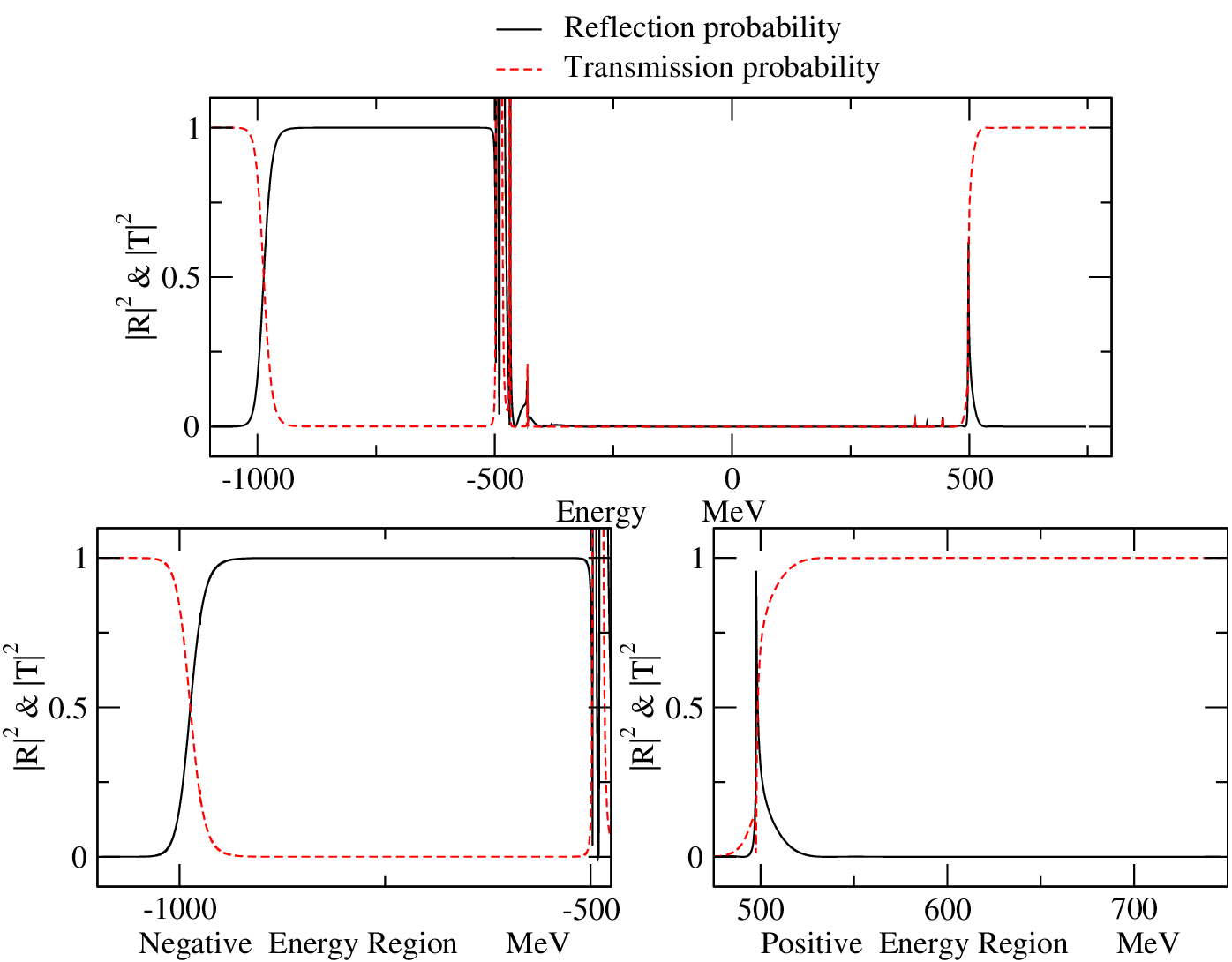}
\caption{In case of SS, the plots of the reflection $R$ and transmission $T$ probabilities versus incident particle energies  which has rest mass energy $m_0c^2=497.648\,\mathrm{MeV}$. The red dashed lines are the transmission, while the black solid ones are the reflection probabilities.  The potential parameters are chosen arbitrarily as $V_0=\frac{3m_0c^2}{2}$, $W=\frac{3m_0c^2}{2}$, $a=1\,\mathrm{fm}^{-1}$,  $L=6\,\mathrm{fm}$. In the first line we see the KG forbidden gap. In the second line, the particle scattering graphs are given in more details. }
\label{SSC_V0_1,5mc^2-W0_1,5mc^2}       
\end{figure*}

\newpage
\begin{figure*}
  \includegraphics[width=1.0\textwidth]{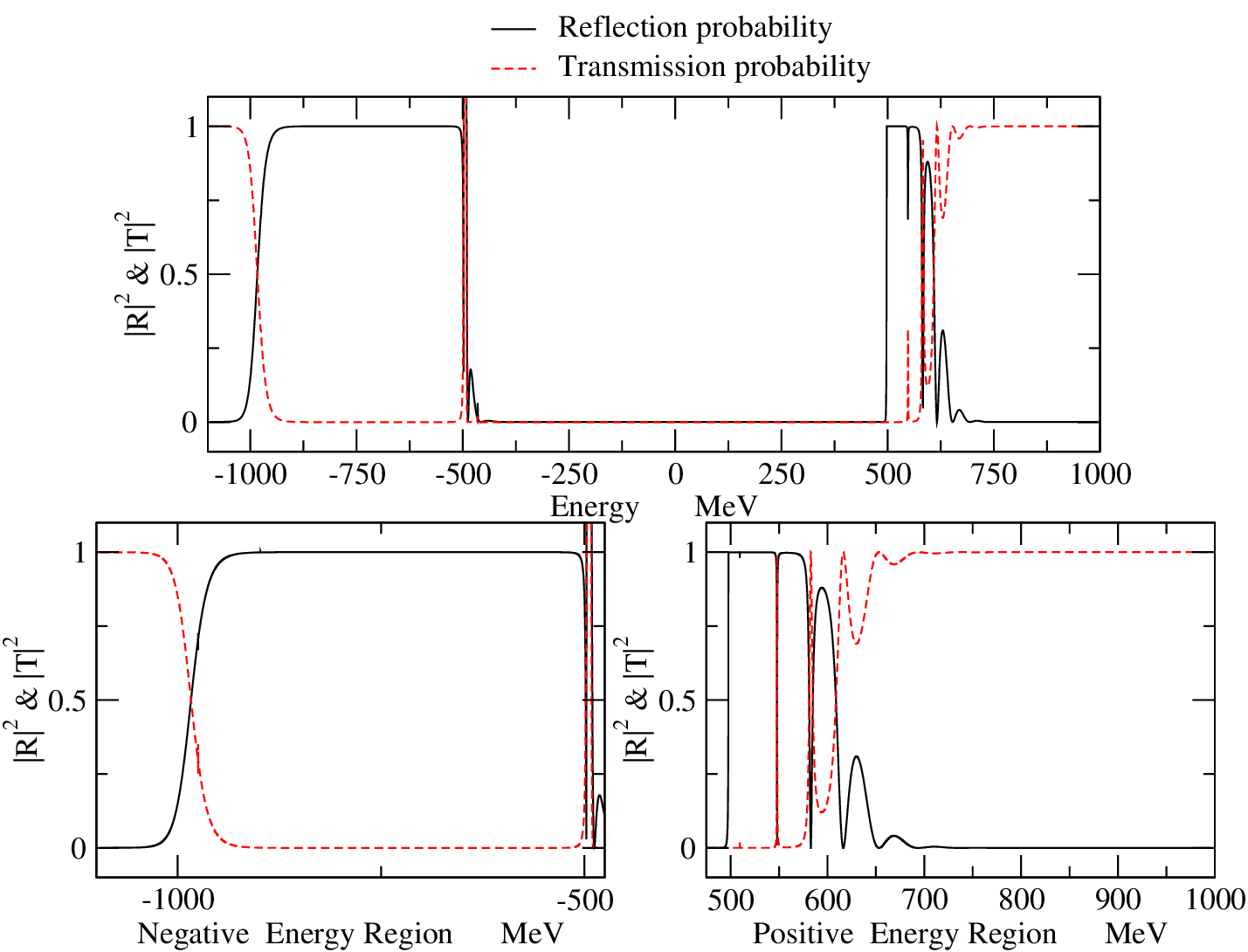}
\caption{In case of SS, the plots of the reflection $R$ and transmission $T$ probabilities versus incident particle energies  which has rest mass energy $m_0c^2=497.648\,\mathrm{MeV}$. The red dashed lines are the transmission, while the black solid ones are the reflection probabilities.  The potential parameters are chosen arbitrarily as $V_0=\frac{3m_0c^2}{2}$, $W=\frac{5m_0c^2}{2}$, $a=1\,\mathrm{fm}^{-1}$,  $L=6\,\mathrm{fm}$. In the first line we see the KG forbidden gap. In the second line, the particle scattering graphs are given in more details. }
\label{SSC_V0_1,5mc^2-W0_2,5mc^2}       
\end{figure*}

\begin{figure*}
  \includegraphics[width=1.0\textwidth]{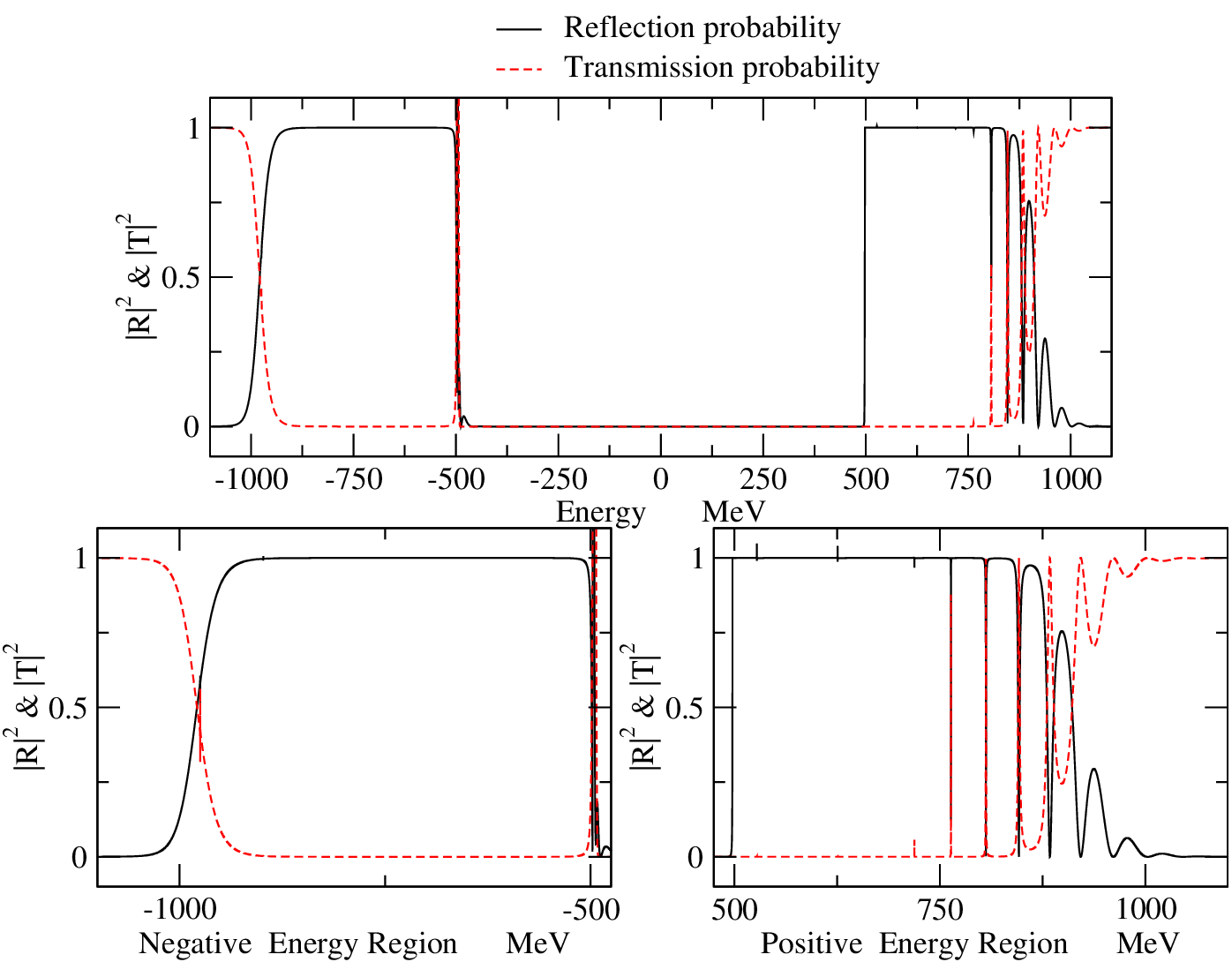}
\caption{In case of SS, the plots of the reflection $R$ and transmission $T$ probabilities versus incident particle energies  which has rest mass energy $m_0c^2=497.648\,\mathrm{MeV}$. The red dashed lines are the transmission, while the black solid ones are the reflection probabilities.  The potential parameters are chosen arbitrarily as $V_0=\frac{3m_0c^2}{2}$, $W=4m_0c^2$, $a=1\,\mathrm{fm}^{-1}$,  $L=6\,\mathrm{fm}$. In the first line we see the KG forbidden gap. In the second line, the particle scattering graphs are given in more details. }
\label{SSC_V0_1,5mc^2-W0_4,0mc^2}       
\end{figure*}

\newpage
\begin{figure*}
  \includegraphics[width=1.0\textwidth]{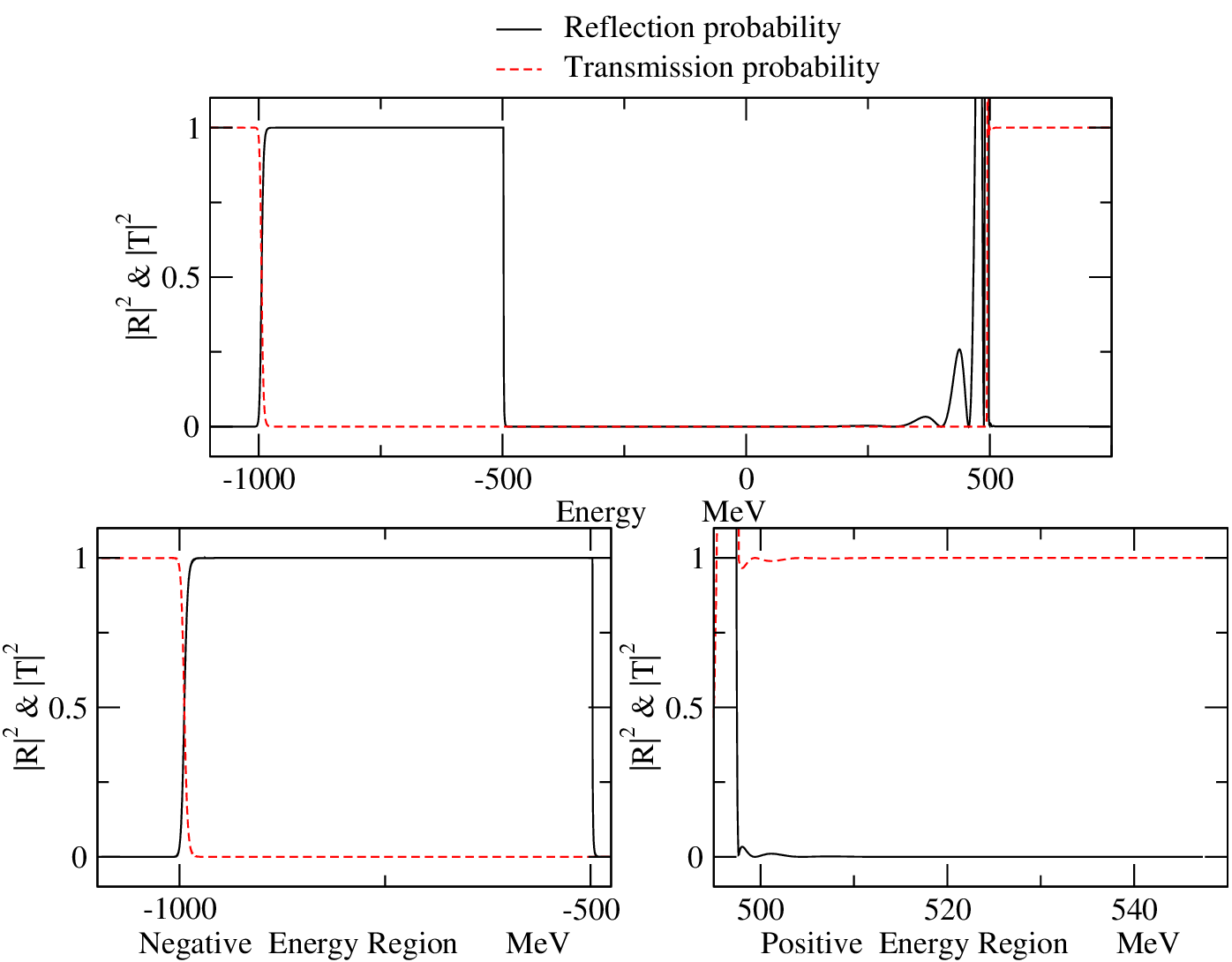}
\caption{In case of PSS, the plots of the reflection $R$ and transmission $T$ probabilities versus incident particle energies  which has rest mass energy $m_0c^2=497.648\,\mathrm{MeV}$. The red dashed lines are the transmission, while the black solid ones are the reflection probabilities.  The potential parameters are chosen arbitrarily as $V_0=\frac{m_0c^2}{2}$, $W=0$, $a=1\,\mathrm{fm}^{-1}$,  $L=6\,\mathrm{fm}$. In the first line we see the KG forbidden gap. In the second line, the particle scattering graphs are given in more details. }
\label{PSSC_V0_0,5mc^2-W0_0,0mc^2}       
\end{figure*}

\newpage
\begin{figure*}
  \includegraphics[width=1.0\textwidth]{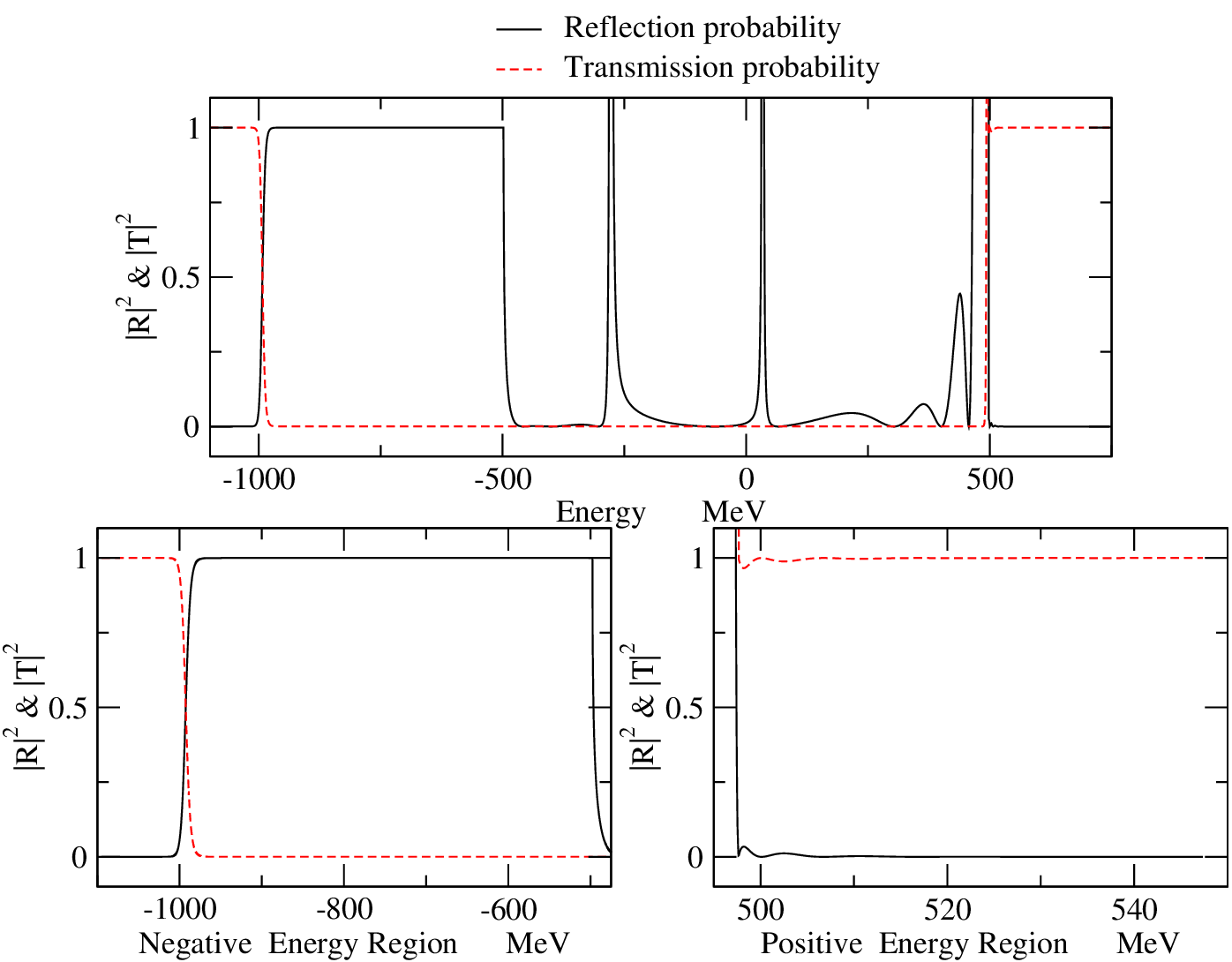}
\caption{In case of PSS, the plots of the reflection $R$ and transmission $T$ probabilities versus incident particle energies  which has rest mass energy $m_0c^2=497.648\,\mathrm{MeV}$. The red dashed lines are the transmission, while the black solid ones are the reflection probabilities.  The potential parameters are chosen arbitrarily as $V_0=\frac{m_0c^2}{2}$, $W=\frac{m_0c^2}{2}$, $a=1\,\mathrm{fm}^{-1}$,  $L=6\,\mathrm{fm}$. In the first line we see the KG forbidden gap. In the second line, the particle scattering graphs are given in more details. }
\label{PSSC_V0_0,5mc^2-W0_0,5mc^2}       
\end{figure*}

\newpage
\begin{figure*}
  \includegraphics[width=1.0\textwidth]{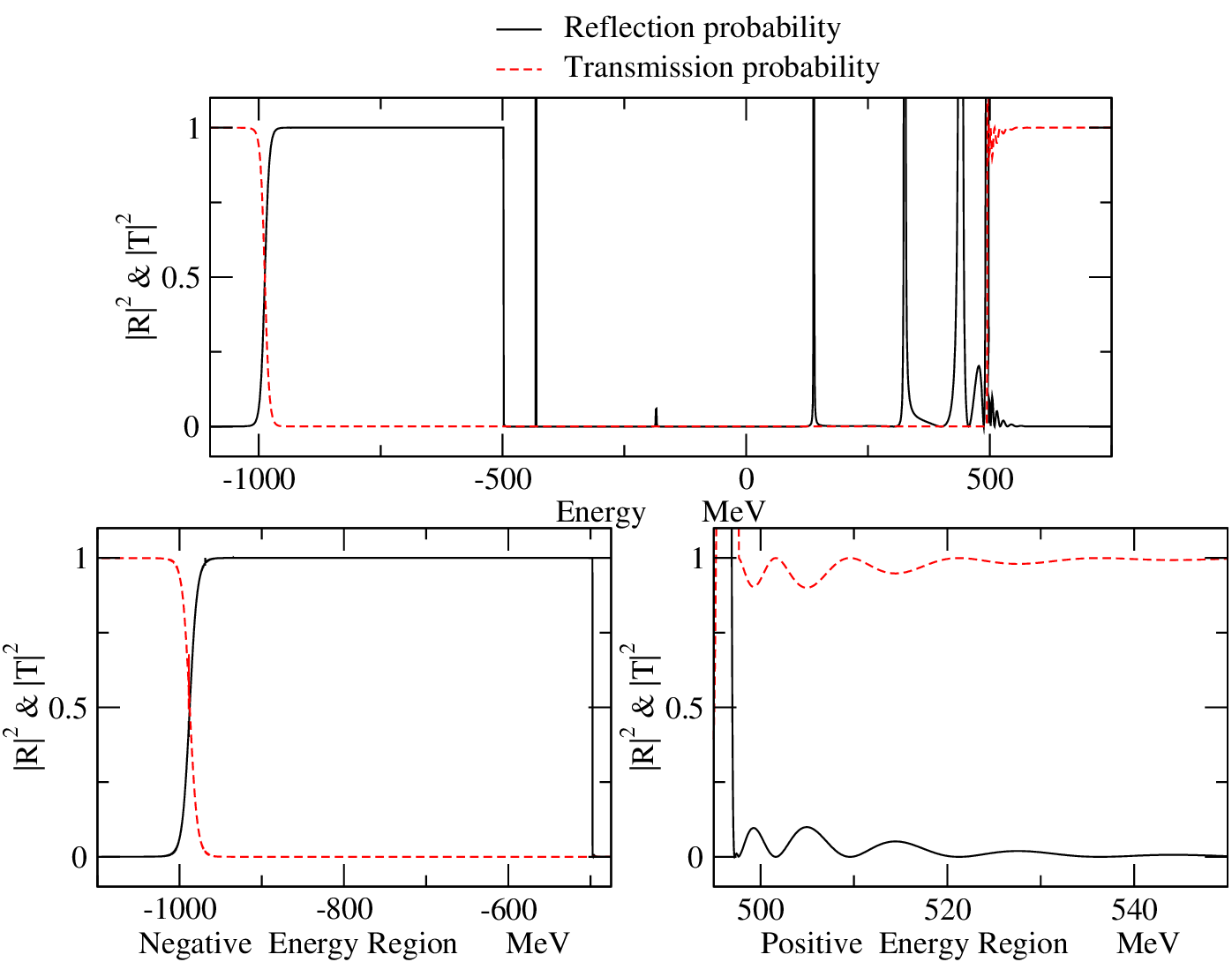}
\caption{In case of PSS, the plots of the reflection $R$ and transmission $T$ probabilities versus incident particle energies  which has rest mass energy $m_0c^2=497.648\,\mathrm{MeV}$. The red dashed lines are the transmission, while the black solid ones are the reflection probabilities.  The potential parameters are chosen arbitrarily as $V_0=\frac{m_0c^2}{2}$, $W= 2m_0c^2$, $a=1\,\mathrm{fm}^{-1}$,  $L=6\,\mathrm{fm}$. In the first line we see the KG forbidden gap. In the second line, the particle scattering graphs are given in more details. }
\label{PSSC_V0_0,5mc^2-W0_2,0mc^2}       
\end{figure*}

\newpage
\begin{figure*}
  \includegraphics[width=1.0\textwidth]{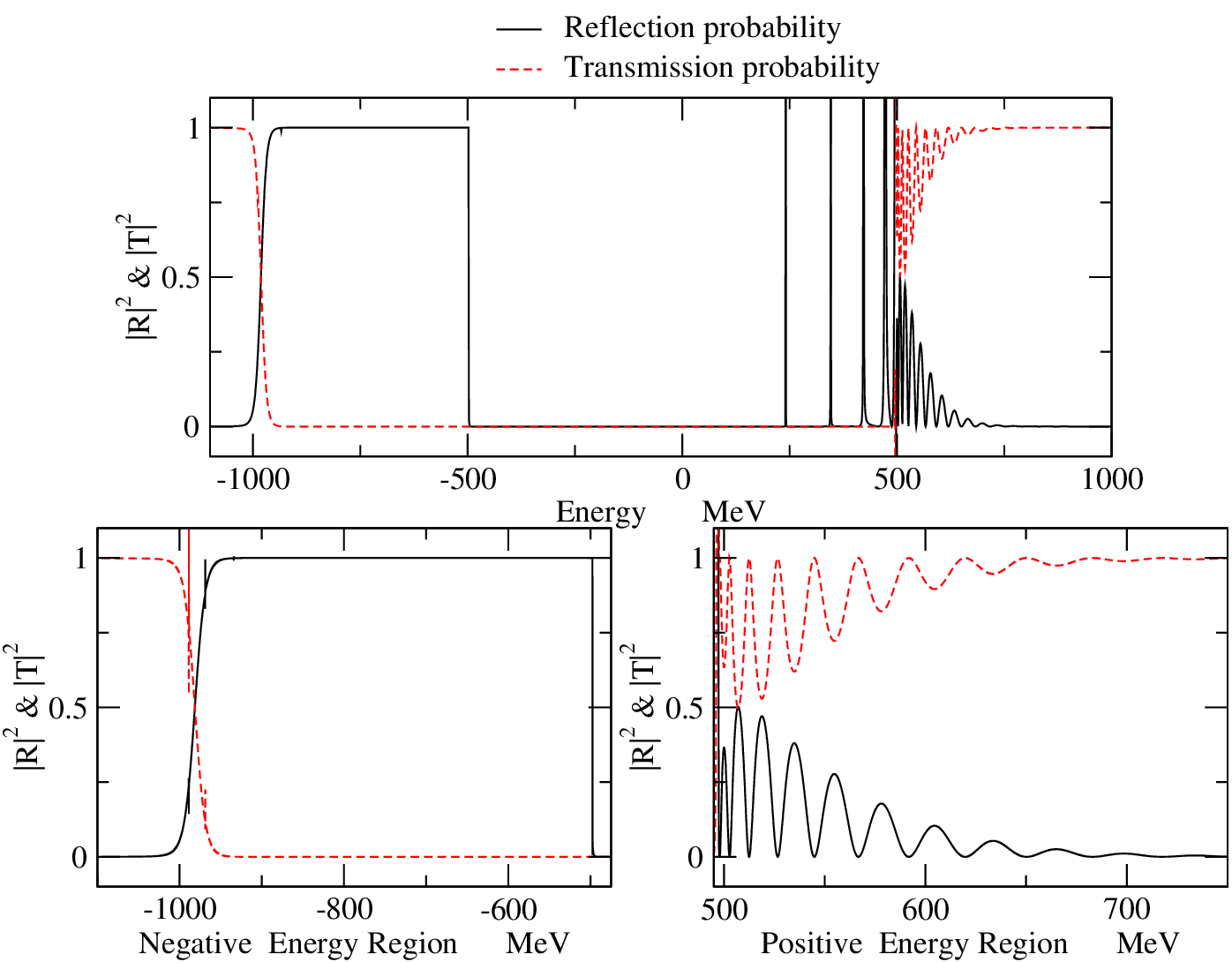}
\caption{In case of PSS, the plots of the reflection $R$ and transmission $T$ probabilities versus incident particle energies  which has rest mass energy $m_0c^2=497.648\,\mathrm{MeV}$. The red dashed lines are the transmission, while the black solid ones are the reflection probabilities.  The potential parameters are chosen arbitrarily as $V_0=\frac{m_0c^2}{2}$, $W= 4m_0c^2$, $a=1\,\mathrm{fm}^{-1}$,  $L=6\,\mathrm{fm}$. In the first line we see the KG forbidden gap. In the second line, the particle scattering graphs are given in more details.}
\label{PSSC_V0_0,5mc^2-W0_4,0mc^2}       
\end{figure*}

\end{document}